\documentclass[12pt,preprint]{aastex}
\usepackage{color}
\usepackage{ulem}

\shorttitle{Evolution of progenitors for ECSNe}
\shortauthors{Takahashi et al.}

\begin{document}

\title{Evolution of progenitors for electron capture supernovae}
\author{Koh Takahashi$^1$, Takashi Yoshida$^2$, and Hideyuki Umeda$^1$}
\affil{
$^1$ Department of Astronomy, The University of Tokyo.\\
$^2$ Yukawa Institute for Theoretical Physics, Kyoto University.\\
Accepted to the ApJ}

\begin{abstract}
We provide progenitor models for electron capture supernovae (ECSNe) with detailed evolutionary calculation. We include minor electron capture nuclei using a large nuclear reaction network with updated reaction rates. For electron capture, the Coulomb correction of rates is treated and the contribution from neutron-rich isotopes is taken into account in each nuclear statistical equilibrium (NSE) composition. We calculate the evolution of the most massive super asymptotic giant branch stars and show that these stars undergo off-center carbon burning and form ONe cores at the center. These cores become heavier up to the critical mass of 1.367 M$_\odot$ and keep contracting even after the initiation of O+Ne deflagration. Inclusion of minor electron capture nuclei causes convective URCA cooling during the contraction phase, but the effect on the progenitor evolution is small. On the other hand, electron capture by neutron-rich isotopes in the NSE region have a more significant effect. We discuss the uniqueness of the critical core mass for ECSNe and the effect of wind mass loss on the plausibility of our models for ECSN progenitors.
\end{abstract}

\section{Introduction}
Electron capture supernova (ECSN) is a distinct class in core collapse supernova (CCSN).
An ECSN progenitor is a super asymptotic giant branch (SAGB) star with a mainly oxygen and neon core, surrounded by a thin helium shell and diffuse hydrogen envelope \citep{nom87}. In an ONe Chandrasekhar mass core, electron capture reactions by $^{24}$Mg and $^{20}$Ne heat the surroundings. As a result, O+Ne burning ignites at the center and generates energy,
and O+Ne deflagration propagates outward. However, the released energy is too small to explode the highly bound core \citep{miy80}. Further electron capture reactions in the central NSE region (Nuclear Statistical Equilibrium) accelerate core contraction. Finally, a proto-neutron star forms and becomes a weak type II supernova \citep{kit06}.

The most distinct point of the progenitor may be its contrasting structure of a highly concentrated core and a diffuse envelope. While the prompt explosion reported in an earlier work of \citet{hil84} was not confirmed in other groups' simulations \citep{bur85, bar87}, a hydrodynamical simulation of collapsing ONe core showed that the delayed explosion powered by neutrino heating takes place even in one-dimensional calculations \citep{may88, kit06}. The successful explosion is found by recent multi-dimensional calculations as well \citep{jan12}, and properties of ECSNe such as nucleosynthesis \citep{wan11, wan13} have been studied.

For observations as well as for theory, a model of ECSN has important implications. Some low luminosity SNe, e.g., SN1997D \citep{tur98}, SN2005cs \citep{pas06,pas09}, can be explained by the explosion model of an ECSN which has a low explosion energy and synthesizes a small amount of $^{56}$Ni. Also observed peculiar compositions in the well-known Crab nebula, such as abundant He and less abundant O, indicate that the Crab supernova SN1054 arose from a collapse of a SAGB star \citep{nom82}. Type IIn SN, which is a SN explosion enshrouded by a dense circumstellar medium, can be explained by an ECSN as well as a CCSN from very massive star that has experienced an intense mass loss phase. Recently, a progenitor of a dust-enshrouded transient SN2008S is found in a pre-explosion image \citep{bot09} and it would have a mass of $\sim$10 M$_\odot$, which is a plausible mass for an ECSN progenitor.

However, there has not been a consistent progenitor calculation from zero age main sequence (ZAMS) to collapse because of the numerical difficulties in calculating the full evolution of SAGB stars. The main difficulties are off-center C burning, thermal pulses, contraction of a highly degenerate core, calculation of electron capture, and propagation of deflagration. These phases have been separately studied by several authors.

The theoretical work on a collapsing ONe core was initiated by Nomoto and collaborators in the 1980's \citep{miy80, miy87, nom84, nom87}. \citet{miy80} investigated effects of electron capture by $^{24}$Mg and $^{20}$Ne and showed that these effects can be summarized as follows: Firstly, reduction of the electron mole fraction induces core contraction. Secondly, reduction of the electron mole fraction reduces the Chandrasekhar mass. Finally, electron capture affects the energy equation endothermically and exothermically. Moreover, \citet{nom87} followed the core evolution after the initiation of O+Ne deflagration using a He star model and provided the progenitor model for an ECSN. Until now, this model was the only one which could be used for an explosion simulation. In the 1990's, non-explosive evolutionary calculation of solar-metal SAGB stars was investigated by Garc\'{i}a-Berro and collaborators \citep{gar94, rit96, gar97, ibe97, rit99}. Non-solar metallicity \citep{gil05, sie07}, as well as detailed physics such as overshooting \citep{gil07} and thermohaline convection \citep{sie09} were considered in recent studies. Off-center carbon burning and thermal pulses, which require expensive calculations, were extensively investigated by \citet{sie06, sie10}. The contraction phase of an ONe core was investigated by recent works as well, concerning the different nuclear reaction rates \citep{has93}, different convective assumptions \citep{gut96}, and different compositions \citep{gut05}. These simulations stopped at the ignition of O burning, and did not model the continuous deflagration phase.

The main purpose of this work is to calculate a progenitor model for an ECSN from a detailed stellar evolutionary simulation. This calculation treats the main sequence phase, which was omitted in \citet{nom87}. It also models the off-center C burning phase, one of the important improvements in the evolutionary theory for SAGB stars. Using a large nuclear reaction network, we include minor isotopes that are synthesized during the C burning phase and additional electron capture reactions by these isotopes are treated with the Coulomb correction. Updated nuclear reaction rates, especially the new electron capture rate for each NSE composition by \citet{juo10}, is taken into account. The increasing ONe core mass is assumed to result from stationary He burning. This enables us to avoid numerical difficulties during the shell He burning phase and to investigate the full evolution of an ONe core.

The critical core mass for ECSNe, past which the ONe core is unstable due to the initiation of electron capture reactions, is considered to be a uniquely determined quantity. The value, calculated by \citet{nom87}, is used for estimates of the initial mass range for ECSNe \citep{sie07,poe08, pum09}. However, the uniqueness on various parameter settings has not been confirmed yet, owing to lack of calculations. Moreover, update of both numerical prescriptions and physical effects such as diffusive convective mixing and the Coulomb correction on electron capture rates possibly affects the value. Investigation on the parameter dependence of the critical core mass for ECSNe is intended in this work as well.

We report stellar evolution of the most massive SAGB stars with solar composition from its main sequence through ONe core contraction and further deflagration phase, in which the central NSE region extends outward. In the next section, the method of calculation and input physics are explained. The predicted evolutionary path from ZAMS phase to the formation of the ONe core is presented in \S\ 3.1. The core contraction including propagation of the deflagration front are explained in \S\ 3.2. Discussions and conclusions are given in \S\ 4.

\section{Methods}
We modify the stellar evolution code in \citet{yos11} (hereafter, referred as Y\&U11) and \citet{ume12} in order to calculate the late phase of ONe core evolution. The mixing length is set to be 1.5 times pressure scale hight. In the following, the main modified points are described.

\subsection{Spacial resolution}
Mesh points are automatically replaced to achieve required resolutions for calculations of stellar structures.
Different conditions are set for different environments as well as different evolutionary phases.
Especially, for the case of propagation of shell C burning, which requires careful treatment of mesh refining,
\begin{enumerate}
\renewcommand{\labelenumi}{(\roman{enumi})}
\item $|\Delta$log$P|$ $<$ 0.05
\item $|\Delta$log$T|$ $<$ 0.1
\item $|\Delta$log$r|$ $<$ 0.1
\item $|\Delta L_r/L_r|$ $<$ 0.1
\item $\Delta M_r/M$ $<$ $10^{-4}$
\end{enumerate}
are taken as the constraints, where $\Delta f$ means the difference of $f$ between two mesh points
and symbols have their usual meanings.

\subsection{Nuclear reaction network}
\begin{figure}
\includegraphics[scale=.30]{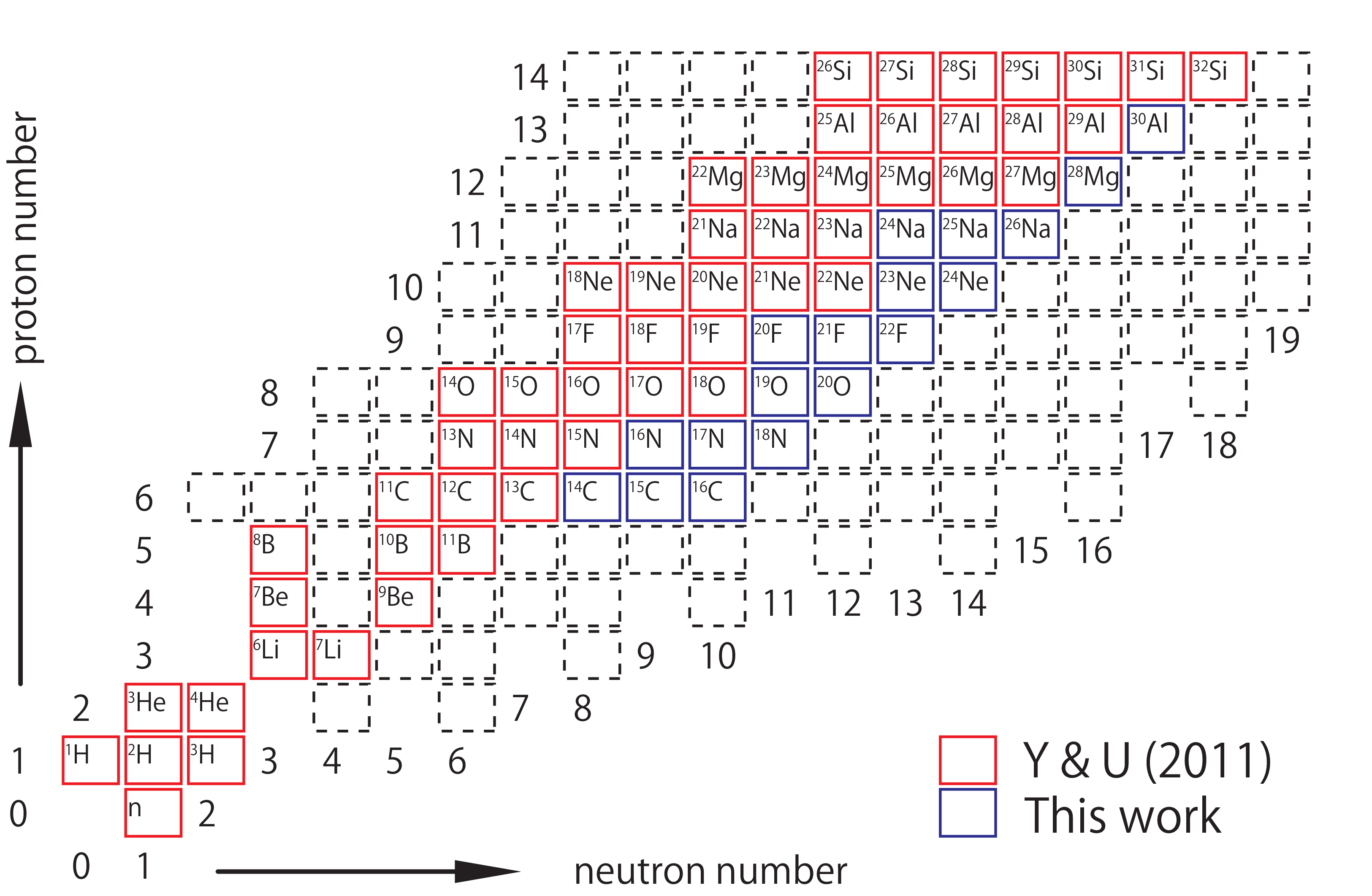}
\caption{isotopes with proton number less than 14 that are included in our calculation. Eighteen blue isotopes are newly included and 300 in total are taken into consideration. Red isotopes had been taken into account in our previous paper (Y\&U11).}
\label{f_added}
\end{figure}

We include 300 isotopes in a reaction network from n, p to Br for calculations of chemical composition evolution and nuclear energy generation. Newly added isotopes from Y\&U11(see also \citet{yos13}) are shown in Fig. \ref{f_added}.

\begin{figure}
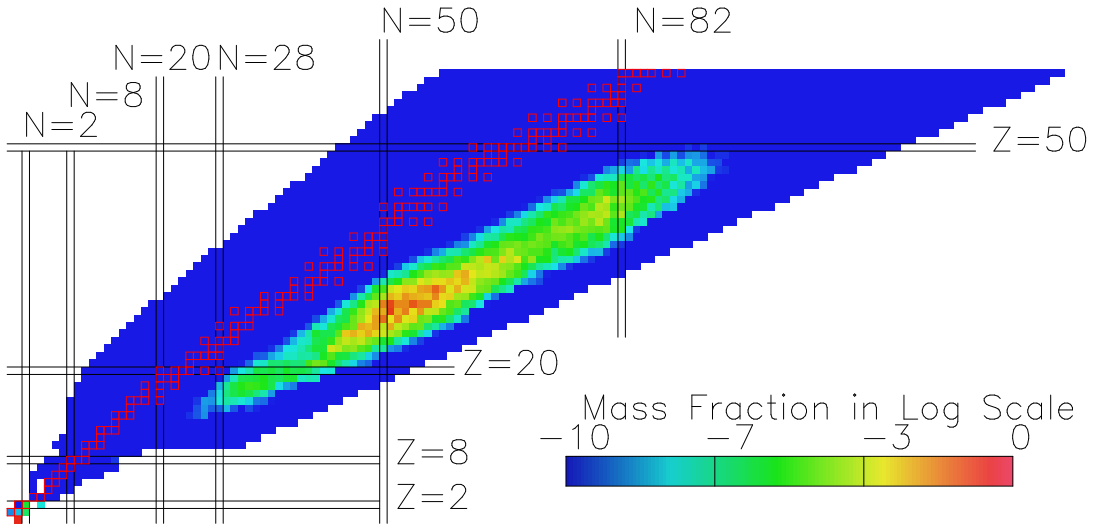

\includegraphics[angle=-90]{FNC_1012.ps}
\includegraphics[angle=-90]{FNC_2801.ps}
\caption{NSE compositions on (N, Z) plane in the cases of (temperature [10$^9$ K], density [g/cm$^3$], electron mole fraction)=(10.0, 10$^9$, 0.49) and (10.0, 10$^{11}$, 0.27). Blue region shows the region of considered 3091 isotopes and red boxes represent stable isotopes. Color contours show the mass fraction distribution.
}
\label{f_nse}
\end{figure}

When the temperature exceeds the value of $5.0\times10^{9}$ K, NSE is assumed to be achieved in the region and the NSE composition is used for calculations of the thermodynamical quantities and nuclear generation rate. In an NSE region in a highly dense ONe core, both density and electron mole fraction vary over wide ranges of (9.0 $\lesssim$ log $\rho$ $\lesssim$ 11.0) and (0.25 $\lesssim$ $Y_e$ $\lesssim$ 0.5). In order to calculate an NSE composition consistently in such a wide parameter space, 3091 isotopes are treated (Fig. \ref{f_nse}). For electron capture reactions in the NSE region, we applied the rate by \citet{juo10} which takes into account roughly 2700 isotopes including very neutron-rich and heavy ones with screening corrections. Though the contributions from positron capture and decay as well as $\beta^-$-decay in the NSE region may affect the stellar evolution, we omit these effects because of the lack of data tables.

\subsection{Convective criterion and diffusive mixing approximation}
In our code, convective mixing is treated as a diffusive process. Mixing beyond the convective boundaries, as in overshooting, is not treated. Convective boundaries are determined by the Schwarzschild criterion, and semi-convective diffusion coefficient given by \citet{spr92} is also applied, thus
\begin{eqnarray}
D_{mix}&=&
\left \{
\begin{array}{ccc}
\frac{1}{3}v_{cv}l_{cv} &\mathrm{for}&
\nabla_{rad}-\nabla_{ad}>\mathrm{min}(0, \frac{\phi}{\delta} \nabla_{\mu})\\
f_{sc}D_{therm}
\frac{\nabla_{rad}-\nabla_{ad}}{\frac{\phi}{\delta}\nabla_{\mu}}
&\mathrm{for}&
0<\nabla_{rad}-\nabla_{ad} \le \frac{\phi}{\delta} \nabla_{\mu},\\
\end{array}
\right.
\end{eqnarray}
where $v_{cv}$ and $l_{cv}$ are convective velocity and scale length determined by the mixing length theory, respectively.
$D_{therm}\equiv \frac{1}{C_p \rho}\frac{4acT^3}{3\kappa\rho}$ is the thermal diffusivity,
and $f_{sc}$ is a free parameter taken as 0.3 according to \citet{ume08}.
In order to take into account the effect of degeneracy of electrons both into the criterion and
the coefficient of convective mixing, we define
\begin{eqnarray}
\frac{\phi}{\delta}\nabla_{\mu} \equiv \frac{1}{\delta}
(\phi_i \nabla_i + \phi_e \nabla_e)
\end{eqnarray}
by extending the work in \citet{kat66}, where
\begin{eqnarray}
\delta \equiv -\frac{\partial \mathrm{ln}\rho}{\partial \mathrm{ln}T},
\phi_i \equiv  \frac{\partial \mathrm{ln}\rho}{\partial \mathrm{ln}\mu_i},
\phi_e \equiv  \frac{\partial \mathrm{ln}\rho}{\partial \mathrm{ln}\mu_e},
\nabla_i \equiv \frac{d \mathrm{ln}\mu_i}{d \mathrm{ln}P},
\nabla_e \equiv \frac{d \mathrm{ln}\mu_e}{d \mathrm{ln}P}.
\end{eqnarray}
Note that thermohaline convection which would take place in a region of
$\frac{\phi}{\delta} \nabla_{\mu}<0$ is not treated for the sake of simplicity.

\subsection{Electron capture and $\beta^-$-decay}
\subsubsection{The energy equation}
In a dense ONe core with density $>10^9$ g/cm$^3$, the energy release by electron capture and $\beta^-$-decay by C burning products becomes important in the energy equation. The divergence of the energy flux $L_r$ in the stellar equation is written as
\begin{eqnarray}
\frac{dL_r}{dM_r}&=&\epsilon_n
- \epsilon_{\nu}
+ \epsilon_g
+ \epsilon_{weak}
+ \epsilon_{mix},\\
\epsilon_g&\equiv&-T\Bigl( \sum_k \frac{ds_k}{dt} \Bigl)
\end{eqnarray}
where $s_k$ is the specific entropy for the k-th particle: nuclei, electrons, and photons (\citet{miy80}, see also \citet{rit99}). $\epsilon_n$ and $\epsilon_{\nu}$ are the nuclear energy generation rate and the neutrino energy loss rate due to processes other than electron capture and $\beta^-$-decay respectively. $\epsilon_{weak}$ is the energy generation rate of electron capture and $\beta^-$-decay, and $\epsilon_{mix}$ represents a cooling term owing to the work by convection \citep{cou75, ibe78}. For radiation, the chemical potential is assumed to vanish, while the one of nuclei is ignored up to the achievement of NSE. For electrons, the chemical potential affects the energy equation through both $\epsilon_{weak}$ and $\epsilon_{mix}$. Here we omit the description of other weak processes such as positron capture and $\beta^+$-decay because of their small effects.

The energy generation rate of both electron capture and $\beta^-$-decay consists of three terms: mass difference, neutrino emission, and chemical potential excluding subatomic energy of relevant particles \citep{miy80}. Since we ignore the chemical potential of nuclei in the energy term, the total rate becomes
\begin{eqnarray}
\epsilon_{weak}=\sum_j(\Delta m_j c^2 \pm \langle E_{\nu, j} \rangle -\mu_e)\frac{dn_{e, weak, j}}{dt} \label{eps_ec}
\end{eqnarray}
where $\Delta m_j$ is the mass difference between parent and daughter nucleus, $\langle E_{\nu, j} \rangle$ is the mean energy of emitted neutrino, $\mu_e$ is the chemical potential of electron, and $\frac{dn_{e, weak, j}}{dt}$ is a time derivative of a specific electron number density owing to the j-th electron capture or $\beta^-$-decay, respectively. The sign of the neutrino energy is taken to be positive for electron capture and is taken to be negative for $\beta^-$-decay. The specific electron number density changes according to
\begin{eqnarray}
\frac{d n_{e, weak, j}}{d t}=\mp \lambda_{j} n_j
\end{eqnarray}
where $n_j$ denotes the specific number density of the j-th nucleus, and the sign of the reaction rate of electron capture or $\beta^-$-decay, $\lambda_{j}$, is taken to be negative for electron capture and is taken to be positive for $\beta^-$-decay, so that the emission of a neutrino should always take some energy away. 

According to \citet{ibe78}, we define
\begin{eqnarray}
\epsilon_{mix}=\frac{\partial \mu_e}{\partial r} F_e(M_r)
\end{eqnarray}
where $F_e(M_r)$ is the flux of electrons and is defined as
\begin{eqnarray}
F_e(M_r)=(4\pi r^2 \rho)D_{mix} \frac{\partial n_e}{\partial M_r}.
\end{eqnarray}
Practically, the electron flux is calculated explicitly in our code, integrating the result of chemical mixing, thus,
\begin{eqnarray}
(4\pi \rho r^2)F_e=\int^{M_r}_{0}\frac{d n_{e, mix}}{d t} dM_r
\end{eqnarray}
where $\frac{dn_{e, mix}}{dt}$ is a time derivative of a specific electron number density owing to chemical mixing. In this expression, the energy loss vanishes at the boundaries of the convective region where the net flow of electrons should be zero.

\subsubsection{Correction for the reaction rate of electron capture and $\beta^-$-decay}
The Coulomb screening for the electron capture rate \citep{cou74, gut96, juo10} is taken into account in our code. Originally, the rate of the electron capture by j-th nucleus $\lambda_{ec, j}$ is written as
\begin{eqnarray}
\lambda_{ec, j}=\frac{1}{\pi^2 \hbar^3} \sum_{states} \int^{\infty}_{\epsilon^0_j}
p_e^2 \sigma_{ec, j} \frac{1}{1+\mathrm{exp}(\frac{\epsilon_e-\mu_e}{kT})} d\epsilon_e
\end{eqnarray}
where $\sigma_{ec, j}$ is the electron capture cross section and $\epsilon^0_j$ is the threshold energy of the reaction \citep{juo10}. This equation implies that $\lambda_{ec, j}$ rapidly increases when $\mu_e$ exceeds $\epsilon^0_j$ since electrons obey Fermi-Dirac statistics. In highly dense regions, the equation of state of nuclei should include the Coulomb correction, and this affects the pressure, the entropy, and also the chemical potential owing to the change of Helmholtz energy. Because the chemical potential depends on a proton number $Z$, change in $Z$ through the electron capture $(A, Z)+e^- \rightarrow (A, Z-1)+\nu_e$ will change the threshold energy of the reaction by the amount
\begin{eqnarray}
\Delta \epsilon^0_{ec} &=& \mu(Z-1) - \mu(Z),\\
\mu(Z)&=&-kT\Bigl(\frac{Z}{\bar{Z}}\Bigl)
\Bigl \{
\Gamma_z
\Bigl[ 0.9 + c_1\Bigl(\frac{\bar{Z}}{Z}\Bigl)^{1/3} + c_2\Bigl(\frac{\bar{Z}}{Z}\Bigl)^{2/3} \Bigl]
+\Bigl[ d_0 + d_1\Bigl(\frac{\bar{Z}}{Z}\Bigl)^{1/3} \Bigl]
\Bigl \}
\end{eqnarray}
where $\Gamma_z=Z^{2/3}\bar{Z}^{4/3}\frac{e^2}{a_i kT}, a_i=(\frac{3}{4\pi n_i})^{1/3}$ is the Coulomb coupling parameter for Z nucleus, $\bar{Z}$ is the mean charge of ions, and the four constants are $c_1=0.2843, c_2=-0.054, d_0=-9/16$, and $d_1=0.460$, respectively \citep{dew73}. Therefore the effective threshold energy becomes
\begin{eqnarray}
\epsilon^0_{ec, eff}&=&\epsilon^0_{ec} + \Delta \epsilon^0_{ec}.
\end{eqnarray}
We assume that the energy distribution function for degenerate electrons does not change its shape for a small difference in electron density. Thus to take into account the effect of the Coulomb correction of the rates, we first evaluate the effective electron density that reproduces the effective chemical potential as the corrected Fermi energy,
\begin{eqnarray}
\epsilon_{F, eff}&=&\epsilon_{F} - \Delta \epsilon^0_{ec}\\
 &\equiv& \mu_{e, eff}((\rho Y_e)_{eff}).
\end{eqnarray}
Then the reaction rate with effective electron density and fixed temperature is applied as the corrected electron capture rate,
\begin{eqnarray}
\lambda_{ec, j, eff} \equiv \lambda_{ec, j}((\rho Y_e)_{eff}, T).
\end{eqnarray}
Since $\Delta \epsilon^0_{ec}$ is positive, this correction increases the effective threshold energy, reduces the effective Fermi energy, and thus reduces the rate of electron capture.

The same correction is also applied to $\beta^-$-decay. In the case of $(A, Z) \rightarrow (A, Z+1)+e^-+\bar{\nu}_e$, change in the threshold $\Delta \epsilon^0_{bd}$ is given as
\begin{eqnarray}
\Delta \epsilon^0_{bd} &=& \mu(Z+1) - \mu(Z)
\end{eqnarray}
which takes a negative value and reduces the threshold energy. For $\beta^-$-decay, the decay rate is correlated to the number of unfilled electron states in which the kinetic energy is between the threshold energy and the chemical potential, i.e., $\mu_e < \epsilon_e < \epsilon^0_{bd, eff}$. Therefore, the decrease of the effective threshold energy reduces the rate of $\beta^-$-decay as well electron capture as mentioned above.

\subsection{An approximate treatment of core growth}
After the completion of C burning in a forming ONe core, we approximate core mass growth by shell He burning in a constant rate in order to avoid some numerical difficulties given below. We assume that the envelope will remain and shell He burning will continue until core collapse. We also treat the core as if it were a single star, and the entropy structure at the edge is assumed to retain geometric similarity through the later evolutionary phases. Since the index of entropy structure is typically expressed by using the homology invariants $V \equiv -\frac{\partial \mathrm{ln}P}{\partial \mathrm{ln}r}$ and $U \equiv \frac{\partial \mathrm{ln}M_r}{\partial \mathrm{ln}r}$ \citep{sug81}, the boundary conditions can be taken as
\begin{eqnarray}
\sigma&=&\sigma_1\mathrm{(const.)},\\
\frac{V}{U}&\equiv&-\frac{\partial \mathrm{ln}P}{\partial \mathrm{ln}M_r}\\
&=&\frac{\partial \sigma}{\partial \mathrm{ln}M_r} \mathrm{(const.)}
\end{eqnarray}
where $\sigma \equiv \frac{m_u}{k}\Sigma_k s_k$ is the specific entropy per baryon in units of the Boltzmann constant, and $\sigma_1$ is the specific entropy at the edge of the core, defined to be a constant. Since the edge structure is extremely steep, this approximation would not affect the later core evolution, especially at the central region.

At the H/He boundary, merging of convective regions, or the dredge-out episode named by \citet{ibe97}, takes place in our calculations (\S\ 3.1.3.). Due to an extended convective region in the helium layer, two convective regions in the hydrogen and helium layers merge together. Some envelope hydrogen is mixed into the base of the helium burning shell, resulting in H burning with significant energy production. As described in \citet{poe08}, this energy production makes numerical convergence difficult, and requires a scheme which can simultaneously solve for mixing and reactions. Our code is not equipped with such a scheme at present. Next, the star enters a thermal pulse phase, if stationary H burning is given as a solution for the the hydrogen mixing problem. In order to treat growth of the core mass in more sophisticated way, a large number of thermal pulses should be calculated. This requires an expensive calculation and a full simulation of the phase is difficult.

The approximation of a constant core mass growth is valid. Because of the large number of pulses, it will be plausible to consider the discrete growths as a time-averaged continuous effect. Also, the relaxation time from the rapid H mixing to the stationary burning is too short to influence error in the core mass growth.

Under these assumptions, three rates, 1.0$\times 10^{-5}$ M$_\odot$/yr, 1.0$\times 10^{-6}$ M$_\odot$/yr, and 1.0$\times 10^{-7}$ M$_\odot$/yr are taken as the core growth rate. The middle one is the most likely rate for core growth from shell He burning. This rate is consistent with the work by Nomoto (1987) in which steady He burning is assumed, and  with recent studies by Siess (2010) and Poelarends et al. (2008) in which the thermal pulse phase is calculated. The results shown in \S\ 3 are cases using this likely rate.

\subsection{Late phase of core evolution}
When the timescale of core evolution becomes shorter than that of convection, the well-known mixing length theory \citep{boh58}, which assumes stationary convection, becomes invalid. In order to determine the temperature gradient in such cases, the time dependent mixing length theory formulated by \citet{unn67} is adopted in our calculation. In this scheme, two time differential equations for convective velocity $v_{cv}$ and temperature fluctuation $\Delta T$ are given as
\begin{eqnarray}
\Bigl(\frac{d }{d t}+\frac{v_{cv}}{l_{cv}/2} \Bigl)v_{cv}&=&\frac{\Delta T}{2 \rho T}
\Bigl(\frac{\partial \mathrm{log}\rho}{\partial \mathrm{log} T}\Bigl)_{P}\frac{d P}{d r},\\
\Bigl(\frac{d }{d t}+\frac{v_{cv}}{l_{cv}/2} \Bigl)\Delta T&=&\frac{v_{cv}}{l_{cv}/2}
\Bigl( \frac{l_{cv}}{2H_P} \Bigl) T (\nabla - \nabla_{\mathrm{ad}})
\end{eqnarray}
where $\nabla - \nabla_{\mathrm{ad}}$ represents the excess of temperature gradient
compared with the adiabatic gradient.
Following \citet{nom84}, we take the length scale of time dependent mixing $l_{cv}$ to be shorter than radial distance.
The convective energy flux $F_{cv}$ is written as
\begin{eqnarray}
F_{cv}=c_P \rho \Delta T v_{cv}
\end{eqnarray}
where $c_P$ is the specific heat at constant pressure.
Then, identical to the mixing length theory, equations of total luminosity
\begin{eqnarray}
L_r&=&L_{rad}+4\pi r^2 F_{cv},\\
L_{rad}&=&\frac{16\pi acGM_r T^4}{3\kappa P}\nabla
\end{eqnarray}
are solved to obtain the temperature gradient.

As the timescale of evolution decreases and becomes comparable with the free-fall timescale,
the assumption of hydrostatic structure becomes invalid.
In this work, an inertia term is included in the equation of motion and 
also in the radiative temperature gradient \citep{heg00} as
\begin{eqnarray}
\frac{dP}{dM_r}&=&-\frac{GM_r}{4\pi r^4}\Bigl[1+\frac{r^2}{GM_r}\frac{\partial^2 r}{\partial t^2}\Bigl],\\
\nabla_{rad}
&=&\frac{3\kappa P L_{r}}{16\pi acGM_rT^4}\Bigl[1+\frac{r^2}{GM_r}\frac{\partial^2 r}{\partial t^2}\Bigl]^{-1}.
\end{eqnarray}

\section{Results}
\begin{figure}
\includegraphics[angle=-90,scale=.50]{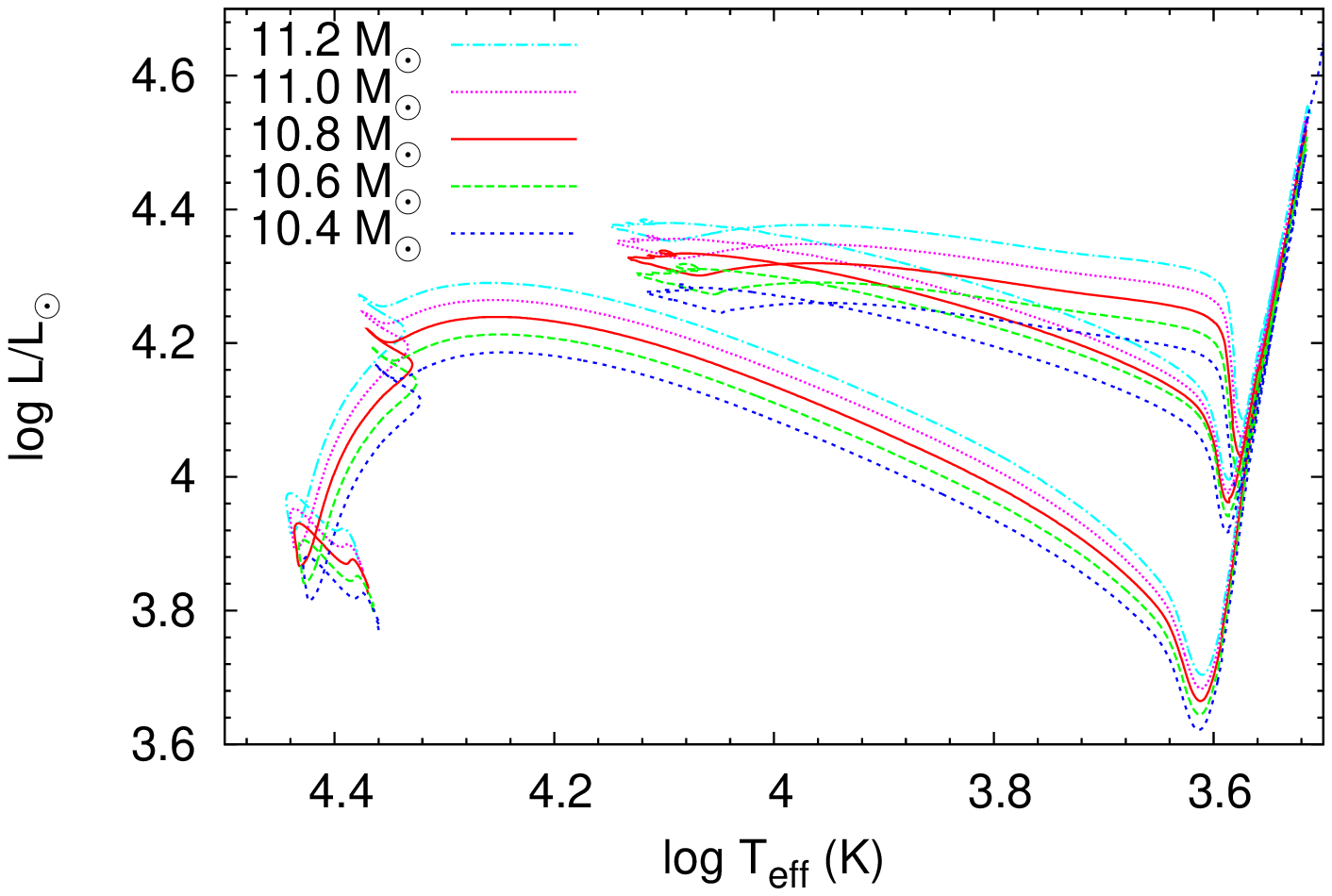}
\caption{Evolution of $\sim$11 M$_\odot$ models in the HR diagram.}
\label{f_HR}
\end{figure}

\begin{figure}
\includegraphics[angle=-90,scale=.50]{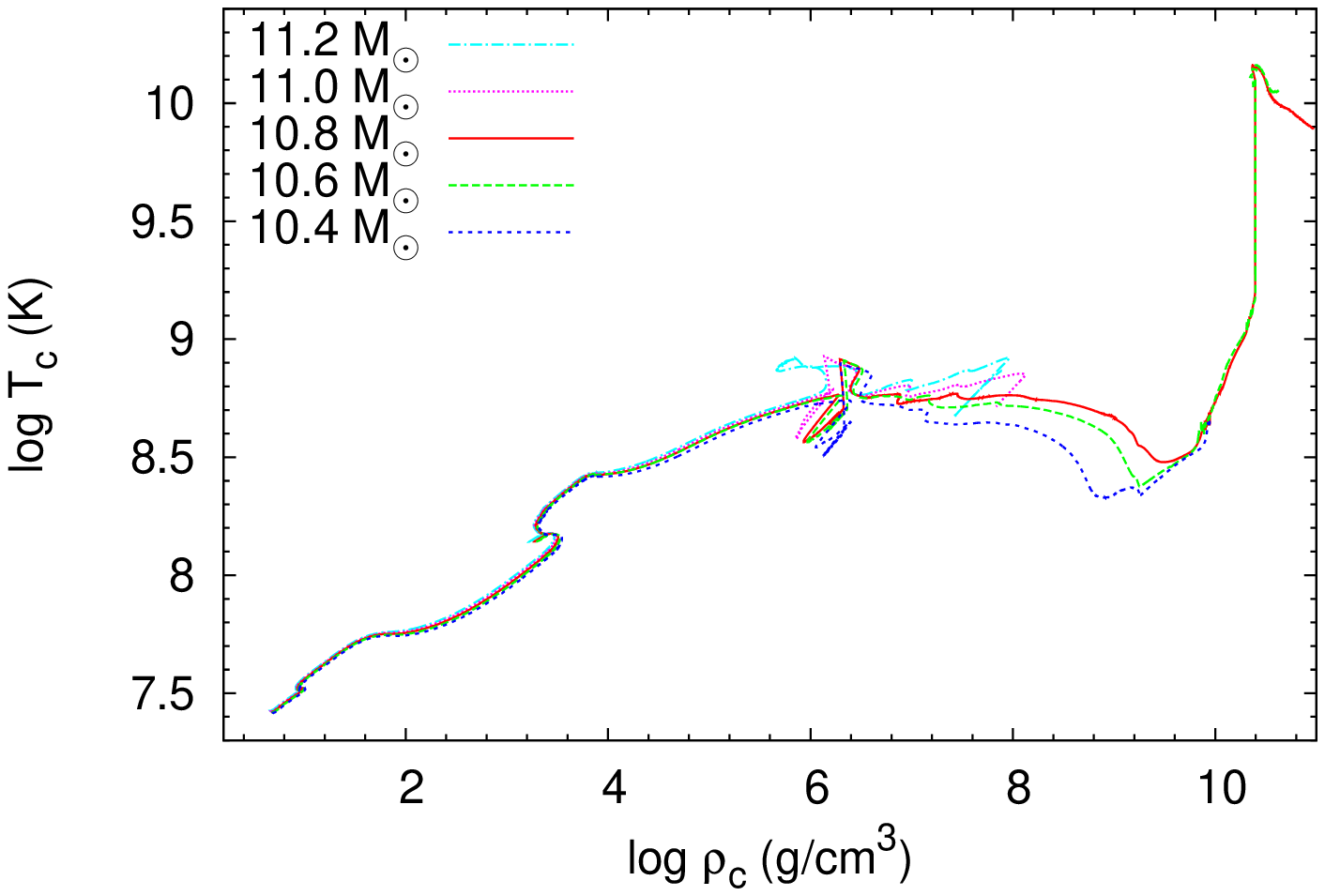}
\caption{Evolution of $\sim$11 M$_\odot$ models in the central density-temperature plane.}
\label{f_rhot}
\end{figure}

We calculated the evolution of 10.4-11.2 M$_\odot$ stars with a metallicity of $Z$=0.02, from ZAMS to O+Ne deflagration for 10.4-10.8 M$_\odot$ models and from ZAMS to off-center Ne ignition for 11.0, 11.2 M$_\odot$ models. Figures \ref{f_HR} and \ref{f_rhot} show the time evolution in the HR diagram, and in the central density-temperature plane, respectively. In Figure \ref{f_rhot}, spikes are shown at $\rho_c \sim 10^8$ g/cm$^3$ for both 11.0 and 11.2 M$_\odot$ models, representing off-center neon ignitions. In our calculation, the minimum initial mass for Ne ignition is 11.0 M$_\odot$ and the CO core mass is 1.35 M$_\odot$. A star with a larger initial mass than this critical mass will form an Fe core and will end up as a normal CCSN \citep{nom88}. Since evolutionary properties of both 10.4 and 10.6 M$_\odot$ models are similar to 10.8 M$_\odot$ model, here we mostly show results of a 10.8 M$_\odot$ model, which has the largest core in these less massive stars and provides the most plausible progenitor model for an ECSN.

The minimum mass for Ne ignition of $\sim11$ M$_\odot$ is large compared with other recent evolutionary calculations \citep{poe08, pum09}. This is because our calculations do not take into account the effect of additional mixing such as overshooting. If an exponentially decreasing diffusion \citep{her00} is considered, our calculation shows that this  minimum mass is reduced by $\sim2$ M$_\odot$ with an overshooting parameter of $f_\mathrm{over}=0.02$. This behavior is quite consistent with the calculation by \citet{sie07}. In fact, inclusion of overshooting significantly alters the relation between initial mass and He core mass. While this severely affects an estimation of an initial mass range for ECSNe, effects on evolutionary results of progenitor calculation are much more mild. Especially, due to a lack of observational constraints, most evolutionary calculations for massive stars only take into account the overshooting before core C burning stages \citep{hir04, lim06}. In this case, our results on the progenitor evolution become fully consistent with these calculations.

\subsection{Pre SAGB evolution}
\begin{figure}
\includegraphics[angle=-90,scale=.50]{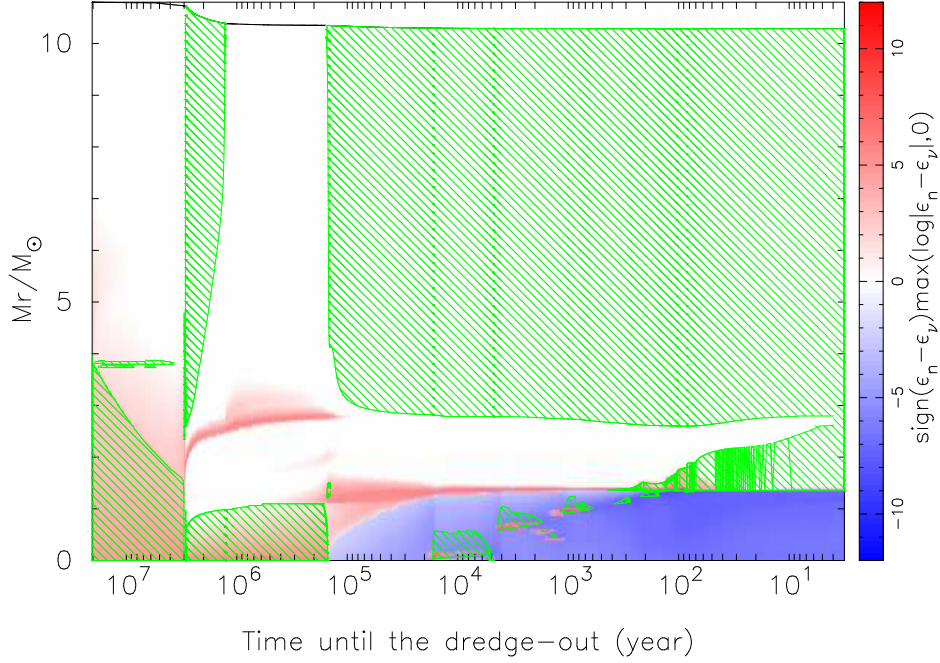}
\caption{The time evolution of convective regions of a 10.8 M$_\odot$ model until completion of the dredge-out. Convective regions are shown in green hatched region in the figure, and log scaled net nuclear energy generation is shown in color.}
\label{f_khd}
\end{figure}

In this section, we summarize the evolutionary results of a 10.8 M$_\odot$ star from its ZAMS phase to the completion of the dredge-out \citep{ibe97}, or merging of convective regions which surround the ONe core. Figure \ref{f_khd} shows the time evolution of convective regions.

\subsubsection{The hydrogen burning stage}
The duration of core H burning is 1.74$\times10^7$ yrs. In this stage, convection develops in the central region owing to the luminosity generated by the CNO-cycle. Core H burning continuously shifts to shell H burning as the central hydrogen burns out, while the H-depleted core contracts and is heated by the release of gravo-thermal energy. The increasing luminosity in the outer shell burning region expands and cools the envelope. As the opacity increases, a convectively unstable region emerges at the surface. The base of the convective region extends inward, and the star becomes a red giant. When the first dredge-up episode occurs, $^{14}$N, the second main product of the CNO-cycle, is dredged up to the surface. Surface composition of CNO isotopes at the end of this episode are summarized in Table \ref{t_surf}.

\subsubsection{The helium burning stage}
When the central density and temperature reach log $\rho_c=3.50$ and log $T_c=8.16$, core He burning takes place. The luminosity of shell H burning decreases, and the base of the convective envelope retreats outward in mass. The entire hydrogen envelope becomes convectively stable, and the star enters a blue-loop on the HR diagram. The core He burning phase continues for 2.76$\times10^6$ yrs, followed by shell He burning after core helium depletion. Since the large luminosity by shell He burning expands and cools the H burning layer, shell H burning dies out. This luminosity re-heats the envelope and induces convection. The star becomes an AGB star, in which a partially-degenerate CO core has formed, surrounded by the He burning shell. At the center of the core, the mass fraction ratio of carbon to oxygen becomes $X(\mathrm{C})/X(\mathrm{O})=0.5728$.

\subsubsection{The carbon burning stage}
\begin{figure}
\includegraphics[angle=-90,scale=.50]{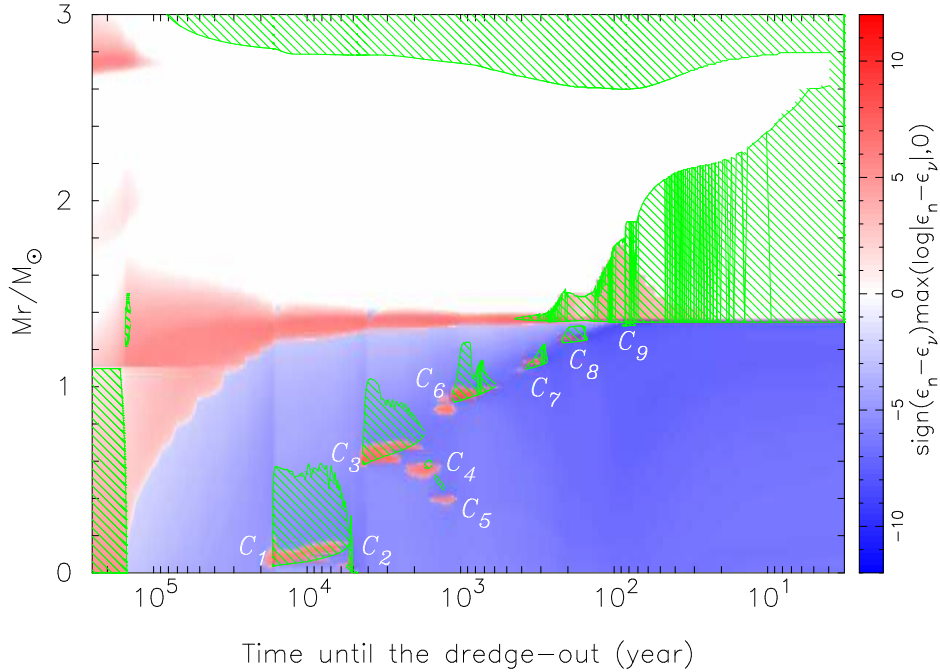}
\caption{Same as Fig. \ref{f_khd} but expanded on the C burning phase.}
\label{f_khd_ex}
\end{figure}

\begin{figure}
\includegraphics[angle=-90,scale=.50]{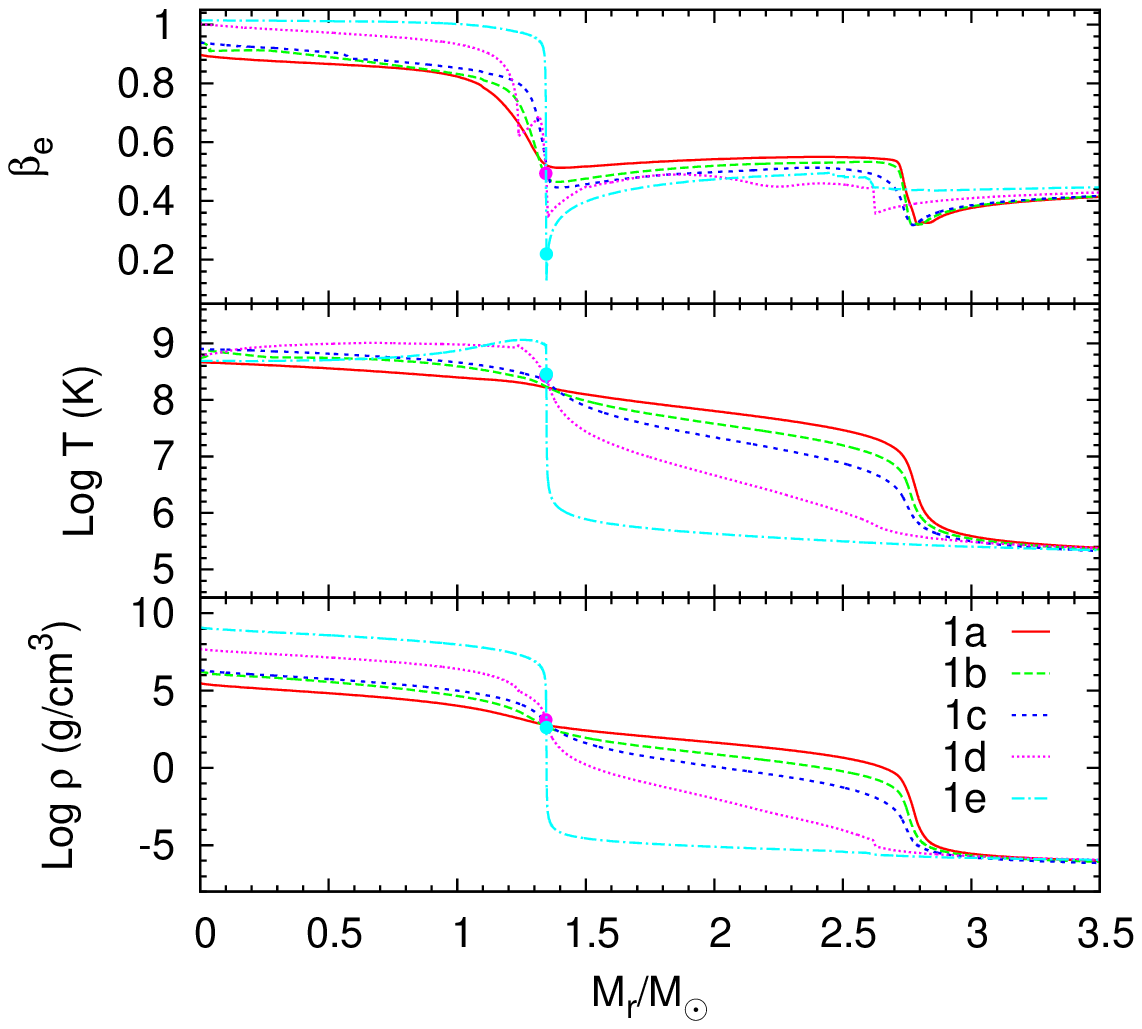}
\caption{Structure evolution in terms of electron pressure fraction (top), temperature (middle), and density (bottom) as functions of mass coordinate at five different stages:
disappearance of convective core He burning (1a),
commencement of the first shell C burning (1b),
ignition at the center of the core (1c),
ignition of the eighth C burning (1d),
and occurrence of the dredge-out(1e).
Points show the locations of the base of the HeBS for (1d) and (1e).
}
\label{f_str}
\end{figure}

In the partially-degenerate CO core, off-center C flashes take place. In our calculation, the first two flashes (C$_{1}$ \& C$_{2}$ in Fig. \ref{f_khd_ex}) arise near the center and the last seven flashes burn outward (from C$_{3}$ to C$_{9}$ in Fig. \ref{f_khd_ex}). These C flashes transform core carbon into neon and other intermediate-mass isotopes. After the end of the sixth C burning, the second dredge-up reduces the mass of the helium layer. These results are consistent with other calculations \citep{sie06, sie07}. Figure \ref{f_str} shows the evolution of the core structure in terms of $\beta_e$, the ratio of electron pressure to the total pressure, temperature, and density at five different stages:
disappearance of convective core He burning (1a),
commencement of the first shell C burning (1b),
ignition at the center of the core (1c),
ignition of the eighth C burning (1d),
and the dredge-out (1e).

In advance of the off-center C ignition, the gravo-thermal heating increases the temperature of the contracting CO core (Fig. \ref{f_str}, 1a). When the central density reaches log $\rho_c=6.28$ and the maximum temperature in the core reaches log $T_{max}=8.81$ at $M_r$ = 0.05 $M_{\odot}$, off-center carbon ignition takes place (Fig. \ref{f_str}, 1b). This is because relatively high temperature activates the cooling process via neutrino emission in the CO core. Neutrino cooling efficiently removes local heat and suppresses temperature increase. The efficiency becomes greater with higher density, so for the inner region of the core, the cooling becomes more effective. The inner region is more degenerate and thus is harder to contract. On the other hand, the still mildly degenerated outer region liberates gravo-thermal energy and increases its temperature, owing to contraction by neutrino cooling. As a result, an inverse temperature gradient appears in a degenerate core.

Since the first C burning expands and cools the central region, the burning front does not propagate inward and the chemical composition at the center does not change. On the other hand, the second shell C burning, which takes place as the first burning dies out, propagates inward and the center of the core ignites at last (Fig. \ref{f_str}, 1c). The surrounding region of the second C burning shell has lower mass fraction of $^{12}$C owing to the first C burning. This weakens the second shell burning, resulting in a smaller expansion of the central region. This enables the second burning flame to propagate inward \citep{sie06}. The propagation is caused by heat conduction from the base of the nearly stationary flame, the mean propagation speed becomes $9.7\times10^{-3}$ cm/sec. Final composition at the center of the core are shown in Table \ref{t_comp} in terms of mass fractions.

\begin{figure}
\includegraphics[scale=.80]{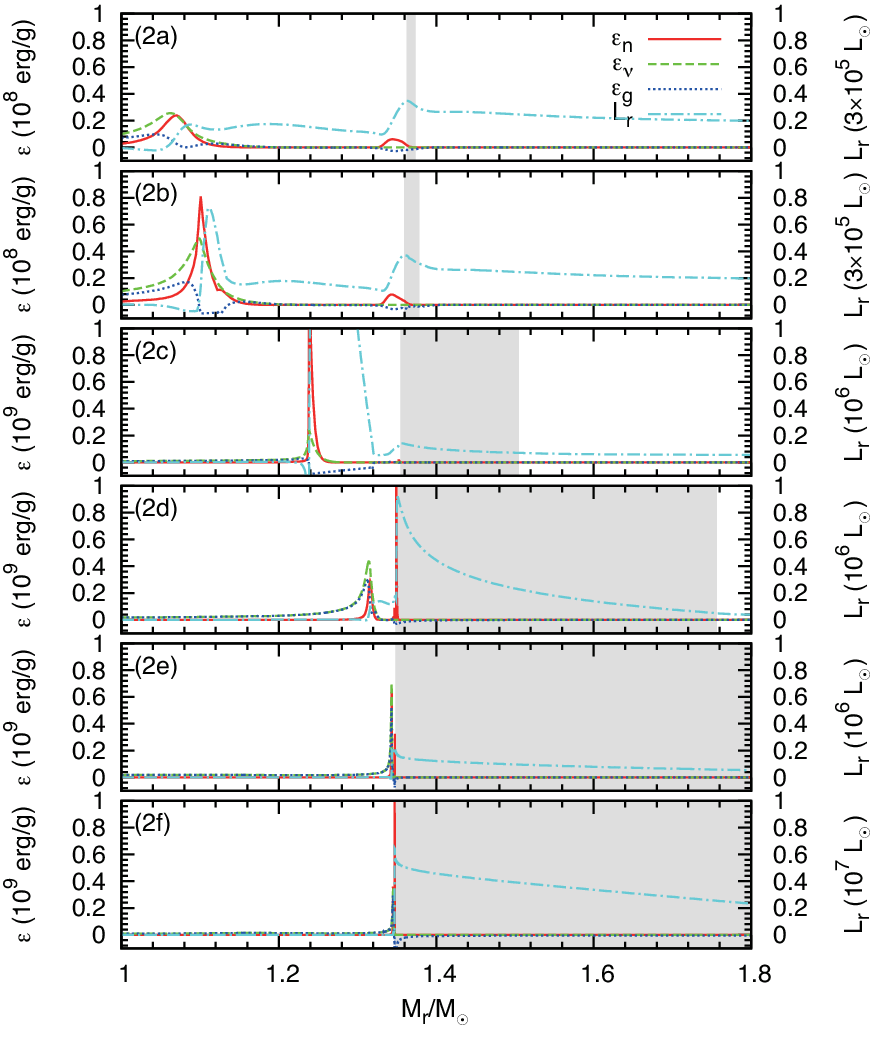}
\caption{Profiles of the energy generation rates and luminosity with mass coordinate at different stages:
occurrence of convection on HeBS (2a),
ignition of the seventh C burning (2b),
ignition of the eighth C burning (2c),
shell He burning at its maximum (2d),
$6.84\times10$ yr after (2d) (2e),
and occurrence of the dredge-out (2f).
Different line types are
the nuclear energy generation (red, solid line),
the neutrino energy loss (green, dashed line)
the gravo-thermal energy release (blue, dotted line),
and the luminosity (cyan, dash-dotted line) respectively.
Gray regions represent convective regions.}
\label{f_edge}
\end{figure}

During the following four C flashes (C$_{3-6}$), the core continues contracting. At the ignition of the seventh C burning, the center of the core is supported only by perfectly degenerate electrons, and the central region clearly shows a temperature inversion (Fig. \ref{f_str}, 1d). On the other hand, radiative pressure has a major fraction of the total pressure both at the flame front ($M_r$=1.24 M$_\odot$) and at the edge of the core ($M_r$=1.35 M$_\odot$). This is due to the high temperature and low density at these regions. Figure \ref{f_edge} shows the evolution of the energy structure and convective region around the edge of the core. Core contraction heats the base of the surrounding He burning shell (HeBS) and increases its luminosity. This He shell burning induces convection at the base of the HeBS (Fig. \ref{f_edge}, 2a), and shell He burning keeps supporting the convection during the seventh and eighth C burning phases (Fig. \ref{f_edge}, 2b \& 2c).

After the end of the eighth C burning, luminosity from the shell He burning takes its maximum value owing to the temperature rise at the base of the HeBS (Fig. \ref{f_edge}, 2d). However, soon the luminosity decreases, and convection in the helium layer is supported by escaping energy from core-edge C burning (Fig. \ref{f_edge}, 2e). The growing convective region in the helium layer merges with the outer convection in the hydrogen envelope. At the moment of the dredge-out episode \citet{ibe97}, the temperature and density steeply drop at the edge of the core from log $T= 9.0\rightarrow6.8$ and log $\rho= 4.8\rightarrow-2.4$ in a narrow mass range of $5\times 10^{-3}$ M$_\odot$ (Fig. \ref{f_str}, 1e). Some envelope hydrogen is mixed into the base of the HeBS. The resulting H burning releases energy at a very high rate of $\sim10^{11}$ erg/sec/g (Fig. \ref{f_edge}, 2f). The ONe core mass, defined by the mass coordinate of maximum energy generation by He burning, is 1.347 M$_\odot$ at the end of the C burning stage. Since the dredge-out episode mixes CNO products with the hydrogen envelope, the surface composition changes after the convective merging (see Table \ref{t_surf}).

\subsection{Evolution of a contracting ONe core}
\begin{figure}
\begin{tabular}{cc}
\includegraphics[angle=-90,scale=.50]{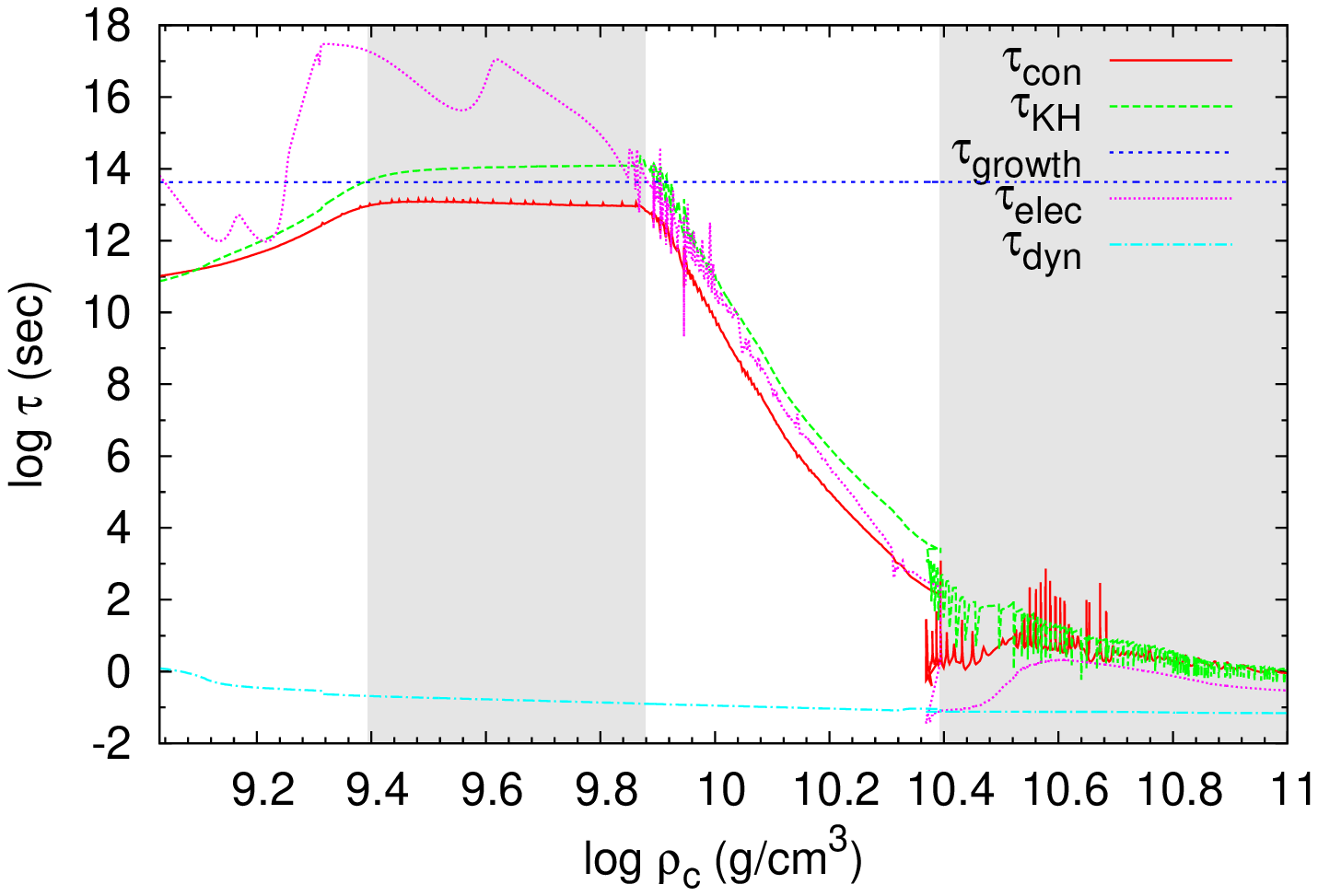}
\end{tabular}
\caption{Evolution of the timescales with density at the stellar center.
Different lines correspond to different timescales:
timescale of core contraction $\tau_\mathrm{con}$ (red, solid line),
Kelvin-Helmholtz timescale $\tau_\mathrm{KH}$ (green, long dashed line),
timescale of core mass growth $\tau_\mathrm{growth}$ (blue, short dashed line),
timescale of electron capture $\tau_\mathrm{elec}$ (magenta, dotted line),
and
dynamical timescale $\tau_\mathrm{dyn}$ (cyan, dash-dotted line).
Background colors shows different sub-phases:
the neutrino cooling phase (log $\rho_c$$\le$9.39),
the core growth phase (9.39$<$log $\rho_c$$\le$9.88),
the electron capture phase (9.88$<$log $\rho_c$$\le$10.39),
and
the deflagration phase (10.39$<$log $\rho_c$).
}
\label{f_time}
\end{figure}

In this section, we show that the evolution of a contracting ONe core can be divided into four sub-phases in terms of driving mechanisms; neutrino cooling, core mass growth, electron capture by $^{24}$Mg and $^{20}$Ne, and O+Ne deflagration. The efficiency of each mechanism is related to evolutionary timescales shown in Fig. \ref{f_time}. The definitions of timescales are given by
\begin{eqnarray}
\tau_{\mathrm{con}}\equiv \frac{d t}{d \mathrm{ln}\rho_c} , \
\tau_{\mathrm{KH}}\equiv \frac{GM_{\mathrm{core}}^2}{R_{\mathrm{core}}(L+L_{\nu})} , \
\tau_{\mathrm{growth}}\equiv \frac{d t}{d \mathrm{ln}M_{\mathrm{core}}} , \
\tau_{\mathrm{elec}}\equiv \frac{d t}{d \mathrm{ln}Y_e} , \
\tau_{\mathrm{dyn}}\equiv \sqrt{\frac{R^3_{\mathrm{core}}}{GM_{\mathrm{core}}}},
\end{eqnarray}
and they represent the timescale of core contraction, the Kelvin-Helmholtz timescale, the timescale of core mass growth, the timescale of electron capture, and the dynamical timescale, respectively.

\subsubsection{Contraction due to neutrino cooling}
From the beginning of contraction until the central density reaches log $\rho_c=9.39$, contraction is caused by neutrino cooling. Figure \ref{f_time} shows that the contraction timescale is close to the Kelvin-Helmholtz timescale, corresponding to the neutrino-cooling timescale. In this phase, the central temperature decreases as the entropy is radiated by thermally activated neutrino emission. As the central temperature decreases, the cooling rate of neutrino emission decreases as well. The timescale of density evolution simultaneously becomes longer.

Although electron captures by $^{27}$Al, $^{25}$Mg and $^{23}$Na proceed in this stage (Fig. \ref{f_enuc}), these reactions have only a minor effect on density evolution. Firstly, this is because the mass fractions of these isotopes are so small that reduction of electron mole fraction does not induce contraction. Secondly, thermal contributions from these reactions are smaller than neutrino cooling. Since the energy production by these electron captures is small, the convective URCA cooling, described in \citet{rit99}, do not affect our calculation.

\subsubsection{Contraction due to core mass growth}
As the core mass increases, both the required pressure to support the core and the actual pressure increase, and thus the central density increases. While the efficiency of neutrino cooling is suppressed by entropy reduction, the efficiency of core mass growth is independent from core structure and becomes constant. After neutrino cooling becomes less effective than core mass growth, the timescale of core growth $\tau_\mathrm{growth}$ starts to limit the contraction timescale.

The stationary core growth forces the central density to increase in a constant rate, and the production rate of gravo-thermal energy becomes constant. Owing to this constant heating, and to less effective neutrino cooling, the central temperature increases proportionally to the central density.

\subsubsection{Contraction due to electron capture}
After the central density exceeds log $\rho_c=9.88$, core contraction is driven by electron capture and thus, $\tau_\mathrm{elec}$ starts to dominate the evolution timescale. The contraction timescale decreases with increasing density owing to the increasing electron capture rate. Soon core growth becomes negligible since $\tau_\mathrm{elec}$ becomes much smaller than $\tau_\mathrm{growth}$. After that, the core mass is frozen and is considered to be at the critical core mass for an ECSN, $M_{\mathrm{EC}}$.

\begin{figure}
\includegraphics[angle=-90,scale=.50]{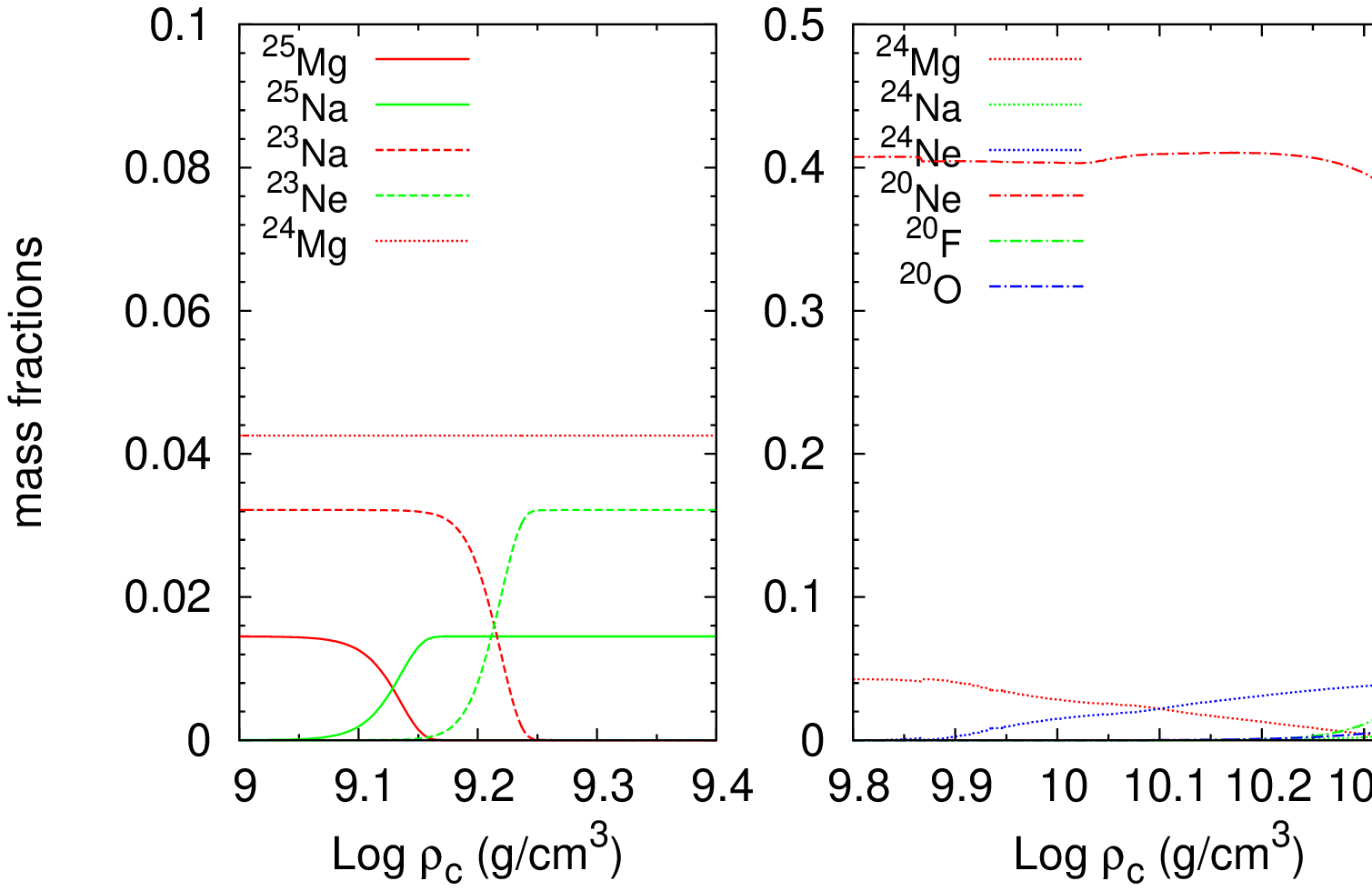}
\caption{Evolution of mass fractions of electron capture nuclei at the stellar center with central density.
For lower density, mass fractions of
$^{25}$Mg, $^{25}$Na (red \& green, solid line),
$^{23}$Na, $^{23}$Ne (red \& green, dashed line), and
$^{24}$Mg (orange, dotted line) are shown.
For higher density,  mass fractions of
$^{24}$Mg, $^{24}$Na, $^{24}$Ne (orange, green, \& blue, dotted line) and
$^{20}$Ne, $^{20}$F, $^{20}$O (orange, green, \& blue, dash-dotted line) are shown.}
\label{f_enuc}
\end{figure}

Figure \ref{f_rhot} shows that the central temperature increases with the central density more steeply in this core growth stage. The increase of temperature is due to heating via electron capture and the main reaction sequence is $^{24}$Mg $\rightarrow$ $^{24}$Na $\rightarrow$ $^{24}$Ne at that time. Electron capture by $^{24}$Mg forms an excited $^{24}$Na$^*$ as a daughter nucleus. When this $^{24}$Na$^*$ decays to the ground state, a $\gamma$-ray results and heats the surroundings. Owing to this additional heating, the electron capture by $^{24}$Mg becomes exothermic. Moreover, both $^{24}$Na$^*$ and $^{24}$Na at the ground state can capture another electron with lower threshold density than with $^{24}$Mg. Therefore, electron capture by $^{24}$Mg results in double-electron capture and releases a large amount of heat. It is noteworthy that the second daughter nucleus $^{24}$Ne becomes the dominant product of electron capture by $^{24}$Mg (Fig. \ref{f_enuc}). As a result, the amount of $\nabla_{\mathrm{rad}}$ becomes sufficiently large compared to $\frac{\phi}{\delta}\nabla_{\mu}$, and the central region is fully mixed by convection. The growing convection supplies fresh $^{24}$Mg to the center where electrons are quickly captured. As a result, the electron mole fraction decreases in the convective region. In accordance with the discussion by \citet{miy80}, convective URCA cooling by $^{24}$Mg-$^{24}$Ne does not take place in this phase, though the reaction drives convection. This is because the threshold density of $\beta^-$-decay for the daughter nucleus $^{24}$Ne, $\sim10^8$ g/cm$^3$, is too small to occur in this phase. While minor electron capture nuclei, such as $^{23}$Na and $^{29}$Si, can induce convective URCA processes, the average energies lost by these reactions are reasonably smaller than the energy released by electron capture by $^{24}$Mg, and thus these processes should be neglected.

After the central density reaches log $\rho_c=10.3$, $^{20}$Ne starts to capture electrons (Fig. \ref{f_enuc}). This electron capture continues driving convection in the same way as electron capture by $^{24}$Mg. The timescale of contraction becomes shorter than the convection timescale, and partial mixing allows a gradient of $X$($^{20}$Ne) to exist in the central region. However, since $^{20}$Ne has a large mass fraction of $\sim$0.4, until the commencement of Ne+O deflagration, the fuel is not consumed and the reaction continues to heat the core.

\subsubsection{O+Ne deflagration}
When the central temperature reaches log $T_c=9.2$, O+Ne burning takes place at the center and the central temperature rises with very short time scale. Since reaction rates of the nuclear burning highly depend on temperature, thermal runaway takes place in the extremely degenerate region. While the temperature steeply increases at the moment of ignition, the variation of pressure and of density become small. When the central temperature exceeds $5\times10^{9}$ K, NSE is assumed to be achieved.

\begin{figure}
\includegraphics[angle=-90,scale=.50]{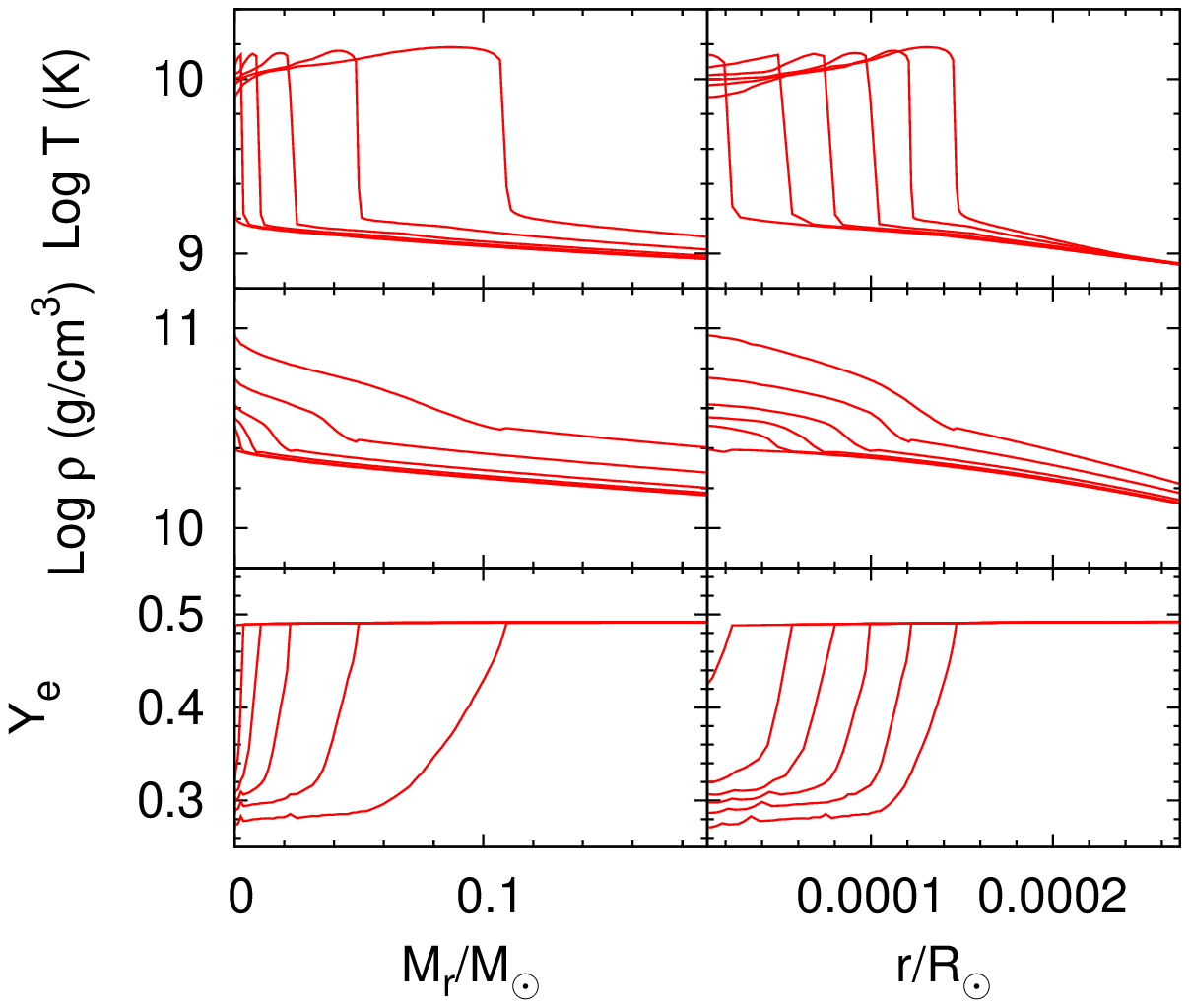}
\caption{Profiles of temperature, density and electron mole fraction during the O+Ne deflagration phase are shown as a function of both mass and radius coordinates.
These profiles are taken at
$1.13\times10^{-2}$,
$5.40\times10^{-2}$,
$9.92\times10^{-2}$,
$1.49\times10^{-1}$,
$2.00\times10^{-1}$, and
$2.34\times10^{-1}$ sec after the ignition at the center of the core.}
\label{f_prop}
\end{figure}

At the boundary of the NSE region and the surrounding ONe region, there must be a negative steep gradient of entropy as a result of the O+Ne burning, and convection must exist. In our calculation, heat transportation by convection is solved (see \S\ 2.6), and the resulting heating at the base of the ONe region becomes much more significant than heating by electron capture by Ne and Mg. Heating increases the temperature, and soon O+Ne burning ignites and NSE is achieved. Top panels of Fig. \ref{f_prop} show the evolution of temperature distribution. The location of the steep temperature gradient is identical to the burning front and propagates outward. The propagation velocity, $1.6\times10^3$ km/sec, becomes smaller than the sound velocity. In other words, the NSE region extends outward in mass by the O+Ne deflagration.

In the NSE region, electron capture reactions by both free-protons and heavy isotopes take place \citep{juo10}. The timescale of electron capture in the NSE region ($\sim$ 0.1 sec) becomes much shorter than in the outer ONe region ($\sim$ $10^2$ sec) (see bottom panels of Fig. \ref{f_prop}). The reduction of the electron mole fraction, coupled with the extension of the NSE region, affects the dynamical evolution of the core. Contrary to the electron capture in the ONe region, the electron capture in the NSE region becomes an endothermic reaction. This is due to the global compositional change in the NSE; unstable neutron rich isotopes are preferred in the lower $Y_e$ environment, and thus reduction of $Y_e$ causes the free energy to be restored in terms of nuclear binding energy. Because of the resulting cooling, a positive entropy gradient appears in the NSE region. The center of the core becomes convectively stable.

The ONe core has been strongly bound by gravity and has a binding energy of $-6.534\times10^{51}$ erg and a total energy of $-5.791\times10^{50}$ erg at the commencement of O+Ne ignition. While energy injection by the O+Ne burning is too small to disrupt the whole star, fast reduction of electrons in the NSE region accelerates core contraction. As a result of both energy injection and electron reduction, the contraction timescale slowly decreases during the deflagration phase. After 2.36$\times10^{-1}$ sec from the ignition at the center of the core, the central density reaches log $\rho_c=11.0$ and the deflagration front reaches $M_\mathrm{NSE}=0.12$ M$_\odot$ and $r_\mathrm{NSE}=1.51\times10^{-4}$ R$_\odot$. The binding energy and the total energy of the core become $-7.739\times10^{51}$ erg and $-9.158\times10^{50}$ erg, respectively.

\subsection{Parameter dependences on $M_\mathrm{EC}$}
\begin{figure}
\includegraphics[angle=-90,scale=.50]{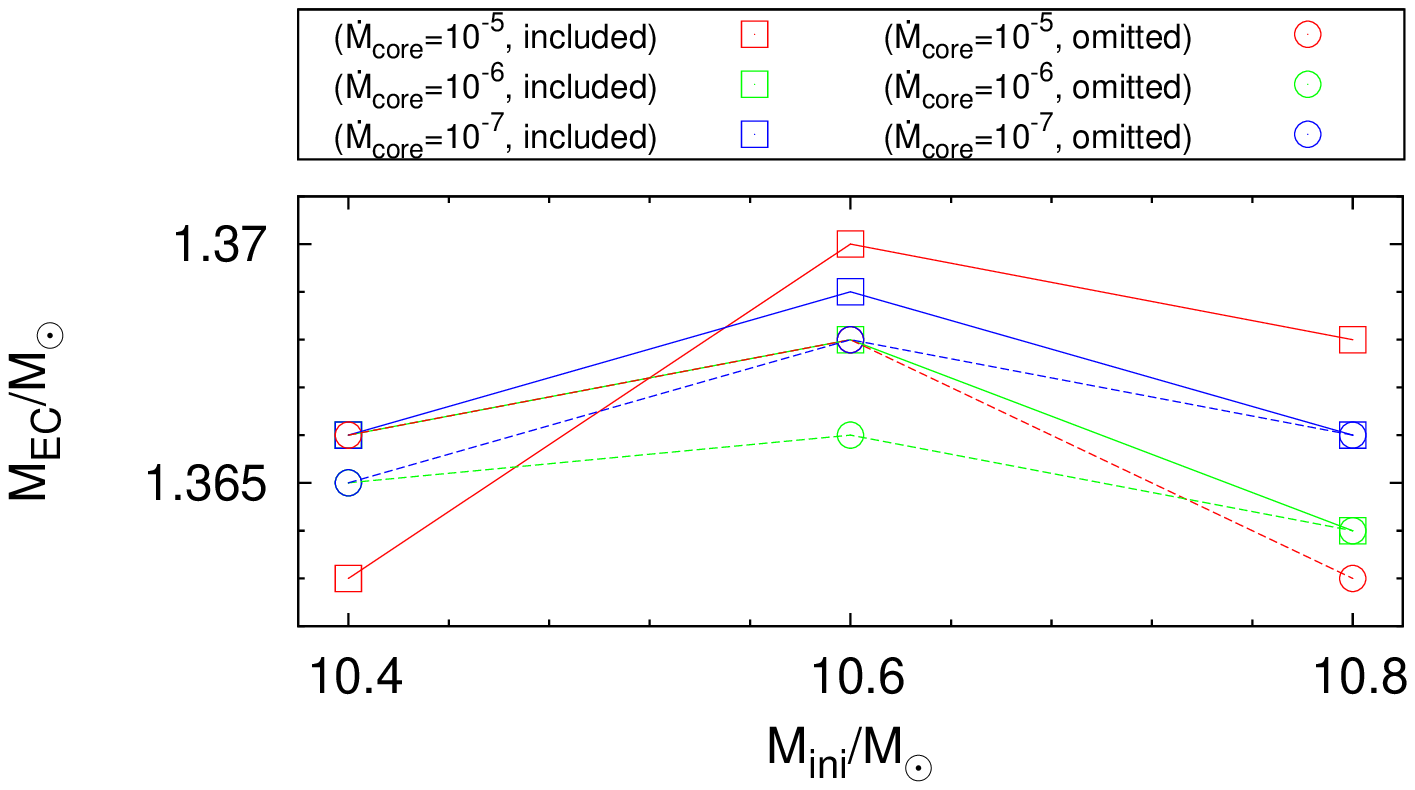}
\caption{The critical core masses for ECSNe calculated with different settings.
Different colors show different core growth rates and different point types show different treatments of the Coulomb correction.}
\label{f_mec}
\end{figure}

In order to investigate the uniqueness of the critical core mass for ECSNe, $M_{\mathrm{EC}}$, in various situations, additional test calculations on ONe core contraction are done with different settings. As an initial condition, ONe cores of 1.346, 1.332, and 1.288 M$_\odot$ are taken to be formed in SAGB stellar models of 10.8, 10.6, and 10.4 M$_\odot$. For each ONe core, the later contraction phase is calculated with three different rates of core mass growth of $1.0\times10^{-5}$, $1.0\times10^{-6}$, and $1.0\times10^{-7}$ M$_\odot$/yr and with or without the Coulomb correction for electron capture. Results are shown in Fig. \ref{f_mec}.

Core masses at the end of calculations were taken as $M_{\mathrm{EC}}$ for each model, however, calculations were stopped for several models owing to numerical difficulties during its electron capture phase. This is due to convective URCA process by minor isotopes such as $^{23}$Na and $^{29}$Si (see \S\ 4.2 for discussion). For these models, we took core masses at the emergence of convection as $M_{\mathrm{EC}}$. Because the contraction timescale steeply decreases after electron capture by $^{24}$Mg starts, and the central convection is driven by this reaction, this procedure would cause small errors in our results. For instance, the core growth after the emergence of central convection for $M_\mathrm{ini}=$10.8 M$_\odot$ with standard settings is 0.002 M$_\odot$. 

The mean value of our results with the Coulomb correction is 1.367 M$_\odot$. We estimate the uncertainty of $M_\mathrm{EC}$ at about $\pm$ 0.005 M$_\odot$, taking into account the error from the determination of $M_\mathrm{EC}$ and the variances of the results. When uncertainties from the C burning phase are considered, the uncertainty should increase in total. We assume that the amount of this uncertainty is as large as that from later stages of evolution. Thus the uncertainty of our result is about $\pm$0.01 M$_\odot$ in total.

\section{Discussions and Conclusions}

\subsection{The critical core mass for ECSNe}
In our calculation, $M_{\mathrm{EC}}$ is 1.367 M$_\odot$ with a relatively small error of $\pm$0.01 M$_\odot$, and is consistent with the result by \citet{nom87}, $M_\mathrm{EC} = 1.375$ M$_\odot$. In a critical Chandrasekhar mass object with given temperature and composition distributions, the central density can be represented only by the core mass, $\rho_c=\rho_c(M_\mathrm{core})$. From different $M_\mathrm{ZAMS}$, different temperature and composition distributions are resulted. The change in $\dot{M}_\mathrm{core}$ results in different durations for neutrino cooling, and leads to different temperature distributions. However, because of the small temperature dependence on degenerate electron pressure, and because of the similar composition of ONe cores for ECSN progenitors from a narrow initial mass range, variations in $M_\mathrm{ZAMS}$ and $\dot{M}_\mathrm{core}$ have little effect on the relation of $\rho_c=\rho_c(M_\mathrm{core})$. Therefore, the critical mass for ECSNe is almost uniquely determined by solving the relation of $\rho_{^{24}\mathrm{Mg}}=\rho_c(M_\mathrm{EC})$, where $\rho_{^{24}\mathrm{Mg}}$ is the threshold density for electron capture by $^{24}$Mg.

Our result shows that inclusion of the Coulomb correction increases $\rho_{^{24}\mathrm{Mg}}$ by $\sim$10\% and increases $M_\mathrm{EC}$ by $\sim$0.003 M$_\odot$ for mean values, but the difference is smaller than the error. Since small variations in $M_\mathrm{EC}$ do not show tendencies on changes in $M_\mathrm{ZAMS}$ and $\dot{M}_\mathrm{core}$, changes in $M_\mathrm{EC}$ due to changes in these parameters would be much smaller than the error. Thus the major component of the $\pm$0.01 M$_\odot$ error should have come from the accumulation of uncertainties in numerical calculations and difficulties in the analysis  described in \S\ 3.3.

\subsection{Convective URCA cooling by minor electron capture nuclei}
When electron capture by $^{24}$Mg drives convection, convective URCA cooling is induced by minor electron capture nuclei such as $^{23}$Na and $^{29}$Si. In a central convective region, these daughter nuclei, which are formed at the center of the core, are mixed with the outer less dense region. Sometimes the density at the outer edge of the convective region is much less than the threshold densities for $\beta^-$-decay of these nuclei. In such cases, reaction rates of $\beta^-$-decay become significantly large, resulting in large energy loss only at the edge of the convection. Some of our models are halted by numerical difficulties caused by this energy loss.

However, this behavior seems unphysical, and slower reactions should occur in reality. Such a large energy loss should immediately stop the convective mixing around the region, and mass fractions of these nuclei are so small that reaction durations should be limited in such a non-convective environment. Therefore, if small fractions of unstable $\beta^-$-decay nuclei are mixed into less dense regions, the resulting endothermic reaction will stop immediately and have only a minor effect on overall evolution. We expect that the convective URCA process during the electron capture phase will not affect later evolution of an ONe core as shown in the 10.8 M$_\odot$ model. To prove this statement, simultaneous solving of the coupling of convective mixing, nuclear reactions, and stellar structure is necessary.

\subsection{Deflagration and core-collapse}
Figure \ref{f_time} shows that the contraction timescale $\tau_\mathrm{con}$ is longer than the dynamical timescale $\tau_\mathrm{dyn}$ even at the end of calculation. Such quasi-static contraction will be altered into dynamical collapse during the deflagration phase, and thus, the core-collapse in an ECSN is caused by electron capture in the extending NSE region after the initiation of O+Ne deflagration, rather than by electron capture by $^{24}$Mg and $^{20}$Ne. In order to provide a plausible progenitor model for an ECSN, the deflagration phase should be considered carefully because propagation of the deflagration front and both energy generation and electron capture in the NSE region can affect the core evolution importantly.

\begin{figure}
\includegraphics[angle=-90,scale=.50]{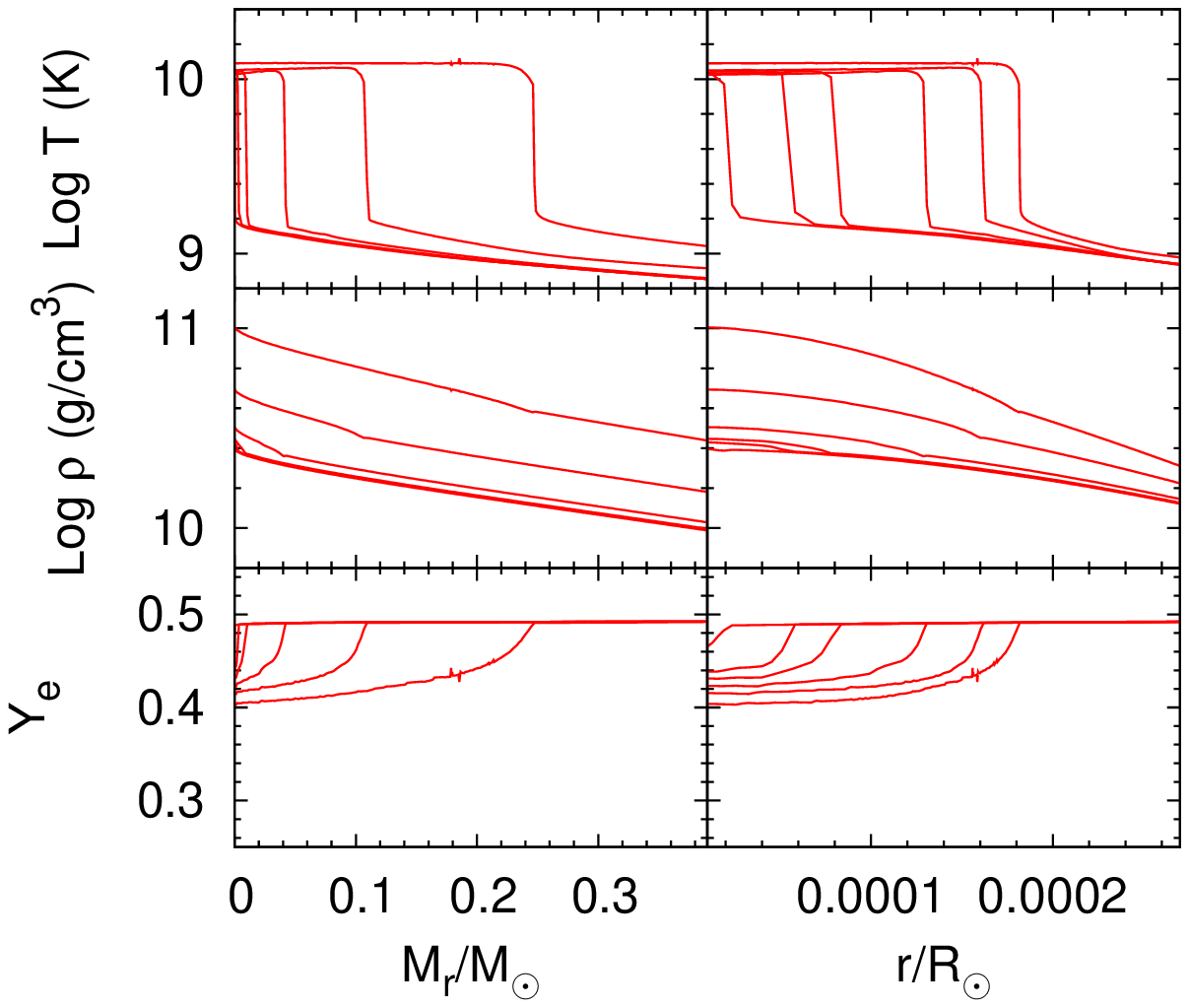}
\caption{Same as Fig. \ref{f_prop} with electron capture rate only by free-protons.
These profiles are taken at
$1.15\times10^{-2}$,
$5.56\times10^{-2}$,
$1.00\times10^{-2}$,
$2.00\times10^{-1}$,
$3.00\times10^{-1}$, and
$3.58\times10^{-1}$ sec after the O+Ne ignition at the center of the core.}
\label{f_prop_proton}
\end{figure}

The obtained deflagration velocity 1.6$\times10^{3}$ km/sec at the end of the calculation ($\rho_c = 10^{11}$ g/cm$^3$) is consistent with the velocity $\sim10^3$ km/sec described in \citet{miy80}. On the other hand, the extension of the NSE region, 0.12 M$_\odot$, is smaller than 0.354 M$_\odot$, obtained by \citet{miy80} and $\sim$0.3 M$_\odot$ by \citet{nom87}. This is due to a different treatment of electron capture in the NSE region. 
Nuclei included in the calculation by \citet{juo10} are much more extended than ones used in \citet{miy80} or \citet{nom87}. At high Ye environment with high proton fraction, electron capture rates are dominated by a contribution from free protons. However, at low Ye environment with extremely low proton fractions, contribution from neutron-rich isotopes becomes dominant and the importance is significant.
In order to confirm the effect of electron capture by neutron-rich isotopes, we calculate the deflagration phase while taking into account only electron capture by free-protons in NSE. The resulting propagation of the deflagration front is shown in Fig. \ref{f_prop_proton}. Comparing with the case of full electron capture, contraction becomes slower owing to a smaller reduction of electrons. Then, the duration of contraction becomes longer and the flame front propagates far from the center by the end of the calculation, even though the flame front has a smaller mean velocity of $8.4\times10^2$ km/sec. The importance of the mechanism of neutrino-heating was suggested in \citet{kit06}. Since the efficiency is affected by the density profile, a central concentrated density profile in our calculation would affect its explosion and the inclusion of electron capture by neutron-rich isotopes will be important.

\subsection{Constraints by mass loss}
It is considered that not all SAGB stars end up as ECSNe because of intense wind mass loss during the SAGB phase. As the core mass increases, a more significant amount of envelope mass will be lost by wind. This limits the duration of the SAGB phase, and if the core mass is less than $M_\mathrm{EC}$ at the end of the SAGB phase, the star ends up as an ONe WD. Therefore the ratio $\xi \equiv \dot{M}_\mathrm{core}/|\dot{M}_\mathrm{env}|$ is important to determine stellar fates.

The minimum $\xi$, with which a star becomes an ECSN just before completely losing its envelope, can be defined as
\begin{eqnarray}
\xi_\mathrm{crit} = \frac{M_\mathrm{EC}-M_\mathrm{core}^\mathrm{ini}}
{M_\mathrm{tot}^\mathrm{ini}-M_\mathrm{EC}}
\end{eqnarray}
where $M_\mathrm{core}^\mathrm{ini}$ and $M_\mathrm{tot}^\mathrm{ini}$ represent an ONe core mass and a total mass at the beginning of the SAGB phase. When an actual $\xi$ for a SAGB star is larger than $\xi_\mathrm{crit}$, the star ends up as an ECSN, and vice versa. If core growth rates are specified by certain simulations, this prescription becomes identical to the definition of the critical mass loss rate $|\dot{M}_\mathrm{env, crit}|$, thus
\begin{eqnarray}
|\dot{M}_\mathrm{env, crit}| \equiv \xi_\mathrm{crit}^{-1} \dot{M}_\mathrm{core},
\end{eqnarray}
and the star with $|\dot{M}_\mathrm{env}| \le |\dot{M}_\mathrm{env, crit}|$ can become an ECSN.

$\xi_\mathrm{crit}$ becomes 2.41$\times$10$^{-3}$, 4.01$\times$10$^{-3}$, and 9.25$\times$10$^{-3}$ for 10.8, 10.6, and 10.4 M$_\odot$ models respectively. Because of large uncertainties for both $\dot{M}_\mathrm{core}$ and $\dot{M}_\mathrm{tot}$, a current estimate of $\xi$ is highly uncertain. Here, according to the results by \citet{poe08}, we limit the range of $\xi$ as 2.54$\times$10$^{-3}$-1.80$\times$10$^{-2}$. This estimate includes wide ranges of mass loss rates, initial masses, and thus mass growth rates, and the minimum value will be too small to be applied to the most massive SAGB stars in our calculation. Even from this crude estimate, our model of a 10.8 M$_\odot$ star is plausible for the progenitor of an ECSN. This conclusion will be robust if the inclusion of overshooting reduces the total mass by $\sim$2 M$_\odot$ and rises $\xi_\mathrm{crit}$ to 3.13$\times$10$^{-3}$.

\subsection{Conclusions}
Stellar evolution for the most massive SAGB stars having solar abundances is calculated in this work, and we provide progenitor models for ECSNe with detailed evolutionary calculation for the first time. In order to avoid the numerical difficulties, and an expensive calculation for thermal pulses, we assume constant core mass growth rate as a result of He shell burning after the completion of core C burning. After this, the core is assumed to retain a geometrically similar edge structure. As to the high temperature region with $> 5.0\times10^9$ K which appears after the initiation of the O+Ne deflagration, NSE is assumed to be achieved.

In such a star, nuclear burning of hydrogen, helium, and carbon take place step by step, then a core mainly made of oxygen and neon forms. The ONe core, which is supported by degenerate electron gas, contracts due to neutrino cooling, core mass growth by the surrounding shell He burning, and reduction of electron mole fraction by electron capture reactions. When the central temperature increases enough to ignite oxygen, O+Ne deflagration takes place and NSE is achieved. Although the O+Ne burning heats the region, the released energy is too small to explode the highly gravitationally bound core. As electron capture reactions by neutron-rich isotopes as well as by free protons accelerate the core contraction, the deflagration front propagates outward. The core will continue to contract up to the formation of a proto-neutron star. The fate will be a weak type II SN.

In our results, the critical mass for ECSNe, $M_\mathrm{EC}$, is 1.367 M$_\odot$ with a relatively small uncertainty of about $\pm0.01$ M$_\odot$. We show the uniqueness of the value under various initial core masses and various core growth rates. The uncertainty comes from the numerical errors and the analytical error as discussed in \S\ 4.1. Inclusion of the Coulomb correction increases $M_\mathrm{EC}$, however, the effect is smaller than the numerical uncertainties.

Inclusion of minor intermediate mass isotopes such as $^{29}$Si, $^{23}$Na, $^{25}$Mg, and $^{27}$Al does not affect the core evolution and thus $M_\mathrm{EC}$. Though the convective URCA process by these isotopes sometimes causes numerical difficulties in an electron capture phase, this process will have only a minor effect on the whole evolution.

Since the ONe core keeps contracting quasi-statically even during the deflagration phase, the core evolution is importantly affected by propagation of the deflagration front and electron capture in the NSE region. We showed that the assumption of electron capture only by free-protons leads to slower contraction and electron capture by neutron-rich isotopes should be incorporated in the model.

Owing to intense wind mass loss during SAGB phase, not all our models may end up as ECSNe. Accurate estimates for the core growth rate and the mass loss rate are difficult and further investigations are still needed on this topic. However, even under the most strict condition taken from \citet{poe08}, our 10.8 M$_\odot$ model is plausible for a progenitor model of an ECSN.

We are grateful to an anonymous referee for many invaluable suggestions. We would like to thank Andrius Juodagalvis, Karlheinz Langanke, and Gabriel Mart\'{i}nez-Pinedo for providing electron capture rates for NSE compositions. We are also grateful to Ken'ichi Nomoto for his useful discussions. We appreciate the careful reading of the manuscript by Hamid Hamidani and Aaron Christopher Bell. This work was supported in part by JSPS KAKENHI Grant Numbers 22540246 and 23540287.

\newpage

\begin{table}
\begin{center}
\caption{Surface mass fractions at three different stages of ZAMS, the 1st dredge-up, and the dredge-out.\label{t_surf}}
\scalebox{0.7}{
\begin{tabular}{c ccc cccc cccc cccc}
\tableline
\tableline
&$^{1}$H&$^{3}$He&$^{4}$He
&$^{12}$C&$^{13}$C&$^{14}$N&$^{15}$N&$^{16}$O\\
\tableline
ZAMS
&7.059$\times10^{-1}$
&8.946$\times10^{-5}$
&2.740$\times10^{-1}$
&3.196$\times10^{-3}$
&3.847$\times10^{-5}$
&1.164$\times10^{-3}$
&4.599$\times10^{-6}$
&1.011$\times10^{-2}$\\
1DUP
&6.778$\times10^{-1}$
&4.566$\times10^{-5}$
&3.021$\times10^{-1}$
&1.876$\times10^{-3}$
&9.926$\times10^{-5}$
&3.704$\times10^{-3}$
&1.869$\times10^{-6}$
&8.893$\times10^{-3}$\\
DOUT
&5.729$\times10^{-1}$
&3.592$\times10^{-5}$
&4.050$\times10^{-1}$
&3.105$\times10^{-3}$
&9.049$\times10^{-5}$
&4.933$\times10^{-3}$
&1.490$\times10^{-6}$
&7.837$\times10^{-3}$\\
\tableline
\tableline
&$^{17}$O&$^{18}$O&$^{21}$Ne
&$^{22}$Ne&$^{23}$Na&$^{25}$Mg&$^{26}$Mg\\
\tableline
ZAMS
&4.097$\times10^{-6}$
&2.284$\times10^{-5}$
&4.350$\times10^{-6}$
&1.372$\times10^{-4}$
&3.519$\times10^{-5}$
&7.131$\times10^{-5}$
&8.178$\times10^{-5}$
&\\
1DUP
&1.168$\times10^{-5}$
&1.504$\times10^{-5}$
&4.344$\times10^{-6}$
&1.137$\times10^{-4}$
&5.969$\times10^{-5}$
&6.554$\times10^{-5}$
&8.776$\times10^{-5}$
&\\
DOUT
&1.064$\times10^{-5}$
&2.246$\times10^{-4}$
&3.906$\times10^{-6}$
&4.634$\times10^{-4}$
&7.974$\times10^{-5}$
&5.717$\times10^{-5}$
&1.032$\times10^{-4}$
&\\
\tableline
\end{tabular}}
\end{center}
\end{table}

\begin{table}
\begin{center}
\caption{Central mass fractions of the fourteen most abundant isotopes at the end of the core carbon burning phase.\label{t_comp}}
\scalebox{0.9}{
\begin{tabular}{ccc ccc c}
\tableline
\tableline
$^{16}$O&$^{20}$Ne&$^{24}$Mg&
$^{23}$Na&$^{25}$Mg&$^{26}$Mg&$^{27}$Al\\
\tableline
4.783$\times10^{-1}$&
4.074$\times10^{-1}$&
4.255$\times10^{-2}$&
3.217$\times10^{-2}$&
1.451$\times10^{-2}$&
7.952$\times10^{-3}$&
7.245$\times10^{-3}$\\
\tableline
\tableline
$^{22}$Ne&$^{28}$Si&$^{12}$C&
$^{21}$Ne&$^{30}$Si&$^{29}$Si&$^{32}$S\\
\tableline
2.836$\times10^{-3}$&
2.330$\times10^{-3}$&
1.277$\times10^{-3}$&
8.384$\times10^{-4}$&
3.621$\times10^{-4}$&
3.362$\times10^{-4}$&
1.463$\times10^{-4}$\\
\tableline
\end{tabular}}
\end{center}
\end{table}

\end{document}